\documentclass[12pt,a4paper]{article}

\usepackage{jheppub} 
\usepackage[swedish,english]{babel}
\usepackage[utf8]{inputenc}
\usepackage[T1]{fontenc} 
\usepackage{graphicx}
\usepackage{amsmath}
\usepackage{amssymb}
\usepackage{amsthm}
\usepackage{latexsym}
\usepackage{color}
\usepackage{bm}
\usepackage{dsfont}
\usepackage{slashed}
\usepackage{hyperref}
\usepackage{breqn}
\usepackage{mathtools}
\usepackage{bm}
\usepackage{cleveref}
\newcommand{\beq}{\begin{equation}}
\newcommand{\eeq}{\end{equation}}
\newcommand{\bea}{\begin{eqnarray}}
\newcommand{\eea}{\end{eqnarray}}

\def\d{\mathrm{d}}

\def\beq{\begin{equation}}
\def\eeq{\end{equation}}
\def\bdm{\begin{displaymath}}
\def\edm{\end{displaymath}}

\def\bea{\begin{eqnarray}}
\def\eea{\end{eqnarray}}

\makeatletter
\newcommand{\pushright}[1]{\ifmeasuring@#1\else\omit\hfill$\displaystyle#1$\fi\ignorespaces}
\newcommand{\pushleft}[1]{\ifmeasuring@#1\else\omit$\displaystyle#1$\hfill\fi\ignorespaces}
\newenvironment{multicases}[1]
  {\let\@ifnextchar\new@ifnextchar
   \left\lbrace%
   \array{@{}l*{#1}{@{\quad}l}@{}}}
  {\endarray\right.\kern-\nulldelimiterspace}
\makeatother

\begin{document}

\title{Holographic Response of Electron Clouds}

\author[a]{U.~Gran,}
\author[a]{M.~Torns\"o}
\author[b]{and T.~Zingg}

\affiliation[a]{Department of Physics,
Division for Theoretical Physics,
Chalmers University of Technology\\
SE-412 96 G\"{o}teborg,
Sweden}
\affiliation[b]{Nordita,
Stockholm University and KTH Royal Institute of Technology\\
Roslagstullsbacken 23,
SE-106 91 Stockholm,
Sweden}

\emailAdd{ulf.gran@chalmers.se}
\emailAdd{marcus.tornso@chalmers.se}
\emailAdd{zingg@nordita.org}

\abstract{In order to make progress towards more realistic models of holographic fermion physics, we use gauge/gravity duality to compute the dispersion relations for quasinormal modes and collective modes for the electron cloud background, i.e.~the non-zero temperature version of the electron star. The results are compared to the corresponding results for the Schwarzschild and Reissner-Nordström black hole backgrounds, and the qualitative differences are highlighted and discussed.}

\maketitle

\tableofcontents

\section{Introduction}
\label{sec:intro}

Holographic duality \cite{Maldacena:1997re,Gubser:1998bc,Witten:1998qj} has proved to be an effective tool to compute the exact response of strongly correlated media, where perturbation theory is not applicable. In particular, electromagnetic properties such as the conductivity and the dielectric function have been studied. The simplest gravitational bulk to describe media with a finite chemical potential is the planar AdS-Reissner-Nordström black hole (AdS-RN). The AdS-RN bulk has two parameters, mass and charge, corresponding to temperature and chemical potential on the boundary, but only one effective dimensionless parameter, the ratio $\mu/T$.

Physically interesting systems often have a temperature that is much smaller than the chemical potential of the system, which is problematic, as the extremal RN black hole is not a stable solution in that parameter regime \cite{Lucietti:2012xr,Aretakis:2011ha,Aretakis:2011hc}. One mode of instability when the temperature is decreased is
that, as the local \emph{bulk} chemical potential increases, fermionic charge carriers could be supported. For low enough temperatures, the black hole charge can decay into these, leaving a density of charge carriers a distance from the remainder of the black hole. 
These bulk configurations, called `electron clouds', have been previously studied \cite{Puletti:2010de,Hartnoll:2010gu}, and are the focus of this paper\footnote{Note that despite the name, the charged fermions in the bulk are not necessarily electrons. Note also that the holographically dual boundary theory generically lacks quasiparticle excitations, so the exact nature of the particles in the bulk is of limited relevance from the perspective of the boundary theory.}. Further turning down the temperature, one eventually arrives at a solution where the event horizon and the inner edge of the could merge into a Lifshitz geometry without any horizon. This configuration has been called `electron star'~\cite{Hartnoll:2009ns}, due to the similarity to neutron stars, but instead built out of charged fermions.

In this paper we expand on previous work~\cite{Aronsson:2017dgf,Aronsson:2018yhi,Gran:2018vdn}, by computing the quasinormal modes~(QNMs), i.e.~the poles of the screened response function $\chi_{sc}$, and the collective modes (CMs), which correspond to the poles of the physical response function $\chi$, in the electron cloud model. The distinction between QNMs and CMs is important since it is the physical response function $\chi$ which is directly accessible in an experimental setup. The screened response of the system, being the sum of the applied external field and the induced polarization, is internal to the system and hence not directly accessible through experiments. In addition, as CMs are the possible oscillations of the system in the absence of {\em external} fields, they correspond to the poles in the physical response function $\chi$.

The paper is structured as follows: In section \ref{sec:model} we briefly review the electron cloud model we will be studying. The results regarding both CMs and QNMs are presented in section \ref{sec:results}, and compared to the results for the simpler Schwarzschild and Reissner-Nordström models. Section \ref{sec:discussion} contains a discussion of the results, and outlines some interesting lines of future research. Since this paper is meant to represent a comprehensive study of both CMs and QNMs for the electron cloud model we have included some amount of details in the appendices.

\section{The Electron Cloud}
\label{sec:model}

We consider the electron cloud model, which consists of Einstein--Maxwell theory coupled to a perfect fluid of charged particles. For more technical details about the action and equation of motion of this model we refer to~\cite{AMORIM1984259,Hartnoll:2009ns,Hartnoll:2010gu,Puletti:2010de}, as well as appendix~\ref{app:action}. The crucial ingredient is the presence of charged fermionic matter that is approximated in the Thomas--Fermi limit, where the density of states is
\begin{equation}
    g(E)\;=\;\beta E \sqrt{(E^2-m^2)}\,,
\end{equation}
for energies above the cloud particle mass $m$, and where $\beta$ is a parameter determining the density of the cloud.\footnote{Also note that the particles in the could are approximated in the zero temperature limit, as finite temperature effects would correspond to $1/N$ effects~\cite{Hartnoll:2010gu,Puletti:2010de}.} This sets the number density and the energy density inside the cloud as
\begin{equation}
    n\;=\;\int_m^{\mu} g(E)\d E\,,\qquad \rho_{fl}\;=\;\int_m^{\mu} E g(E)\d E\,,
\end{equation}
respectively. The pressure is then given by the usual equation of state,
\begin{equation}
    p\;=\; -\rho_{fl} + \mu_{loc} n\,,
\end{equation}
where we have introduced the local chemical potential
\begin{equation}
  \label{eq:mu}
  \mu_{loc}\;=\; u^{\mu}\left( A_{\mu}+\partial_{\mu}\phi\right),
\end{equation}
which is the effective chemical potential potential felt by a particle.

Considering a static stationary background space-time in this setup, there will be three regions to be considered; a pure RN-solution in the IR, an intermediate cloud-solution and another pure RN-solution in the UV (albeit with potentially different parameters than in the IR).
This follows from the fact that the background solution has a local \emph{bulk} chemical potential, given by \eqref{eq:mu}, that is zero at both the horizon and the boundary, with a maximum in between. Where the chemical potential is large enough to support the charged fluid (that is, larger than the mass of the fluid particles), there will be a cloud, located in that region.

\begin{figure}[ht!]
\centering
\includegraphics[width=0.9\linewidth]{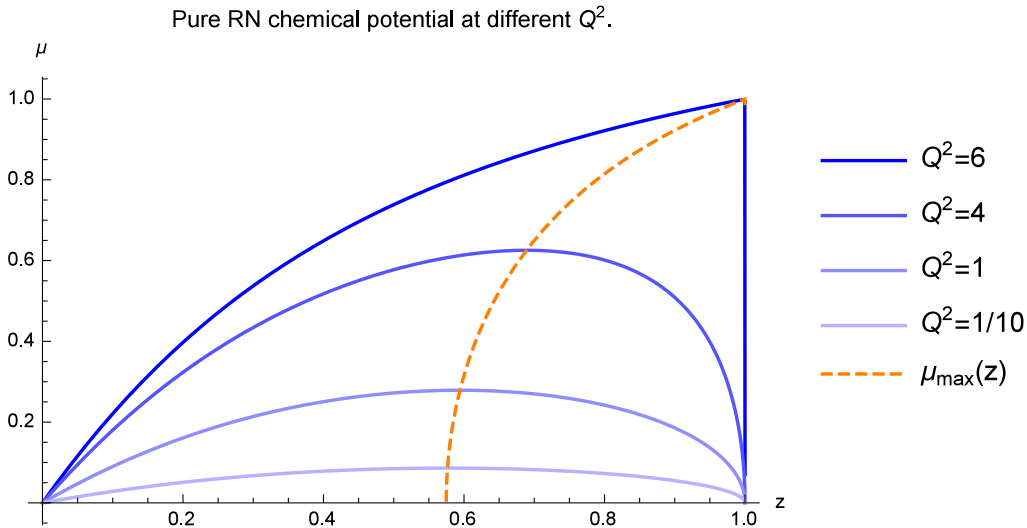}
\caption{Local chemical potential in pure AdS-RN for different values of $Q^2$. The dashed line marks the maximal value, for all different values of $Q^2$.}\label{EC:fig:RNchempot}
\end{figure}

If the charge of the black hole, $Q$, is not large enough, the chemical potential might not be sufficient to support an electron cloud. The solution is then instead a pure planar AdS-Reissner Nordström black hole everywhere. The chemical potential of such models can be seen in figure \ref{EC:fig:RNchempot}, where the horizon is placed at $z=1$, the conformal boundary at $z=0$ and $Q^2=0$ corresponds to the uncharged, Schwarzschild black hole, and $Q^2=6$ corresponds to an extremal black hole. For a cloud particle mass $m$, the system will support a cloud wherever the chemical potential is larger than $m$ (and thus, the solution is no longer pure RN). From the figure we can note that a cloud can only be supported for $m<1$ and the larger the charge $Q$, the closer to the horizon the cloud lies.

\begin{figure}[ht!]
\centering
\includegraphics[width=0.9\linewidth]{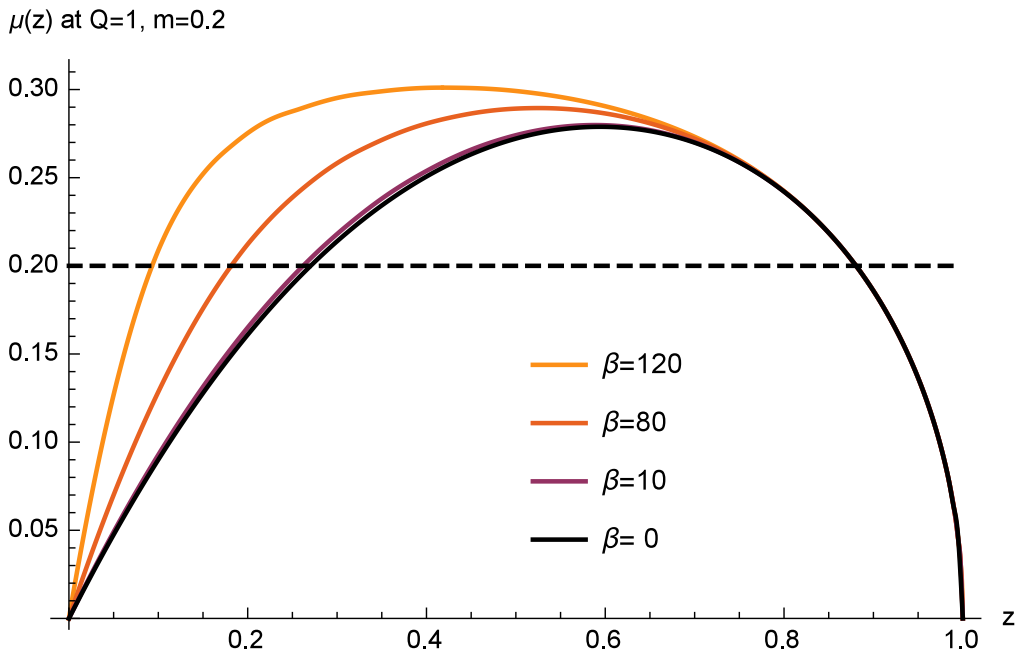}
\caption{The chemical potential in an electron cloud model at $Q=1$ and $m=0.2$ for different values of $\beta$. Note that all chemical potentials are equal only beneath the dashed line ($m$) to the right.}\label{EC:fig:ECchempot}
\end{figure}

Note that for systems with a cloud one also needs to take into account the backreaction of the cloud itself on the region in which it is supported, hence it is only possible to find the inner bound of the cloud from the pure RN-solution in the IR. As the cloud particles have a charge greater than the mass ($m<1$) for stability\footnote{Otherwise the particles would fall into the black hole due to the gravitational pull being larger than the electromagnetic repulsion~\cite{Hartnoll:2009ns}.}, their presence extends the region in which the bulk chemical potential is larger than $m$. How the chemical potential depends on the density parameter $\beta$ is illustrated in figure \ref{EC:fig:ECchempot}. 

\begin{figure}[ht!]
\centering
\includegraphics[width=0.9\linewidth]{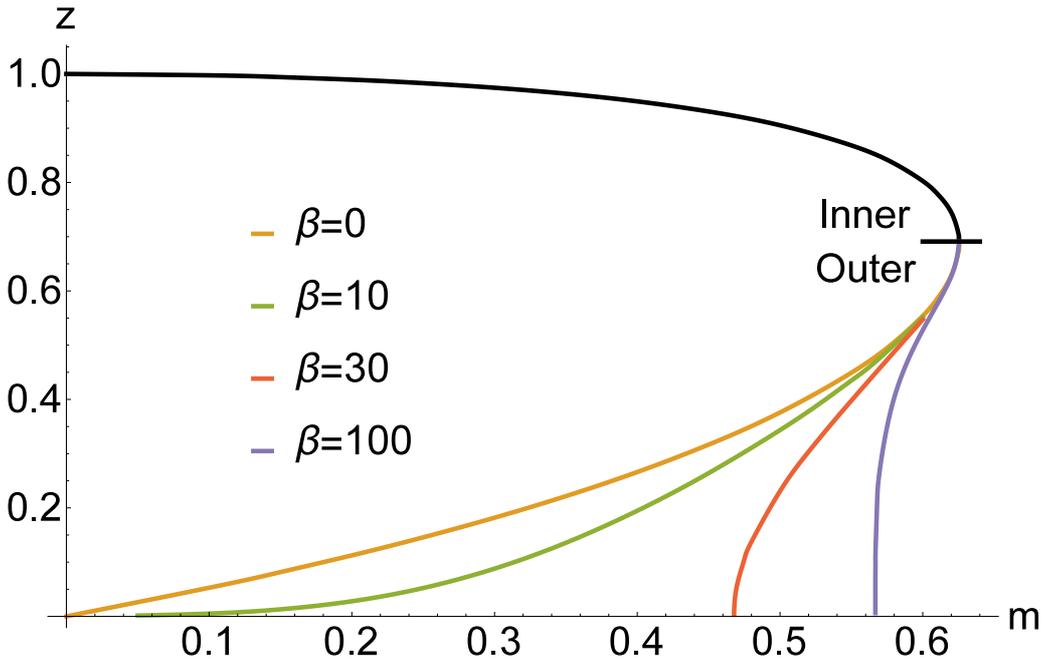}
\caption{The inner (top) and outer (bottom) bounds as a function of mass, measured by the radial coordinate $z$, of the electron cloud for charge $Q^2=4$ and different $\beta$. Note especially that the inner bound is independent of $\beta$ and that for a specific mass, only values of $m>m_{crit}(Q,\beta)$ gives a well defined outer bound.}\label{EC:fig:ECBounds}
\end{figure}

As noted in figure \ref{EC:fig:ECchempot}, the outer bound of the cloud depends non-trivially on the density parameter $\beta$. Similarly, the inner and outer bounds of the cloud depend on the mass parameter, as illustrated in figure \ref{EC:fig:ECBounds}, where for any value of the mass $m$, one can draw a vertical line to find the inner and outer bound of the cloud for some different values of $\beta$, at a fixed $Q^2=4$. Here we can also see that there is a critical limit for large enough $\beta$, since a sufficiently dense electron cloud could source itself indefinitely, indicated by the absence of an outer bound. Whether the outer bound of the cloud extends to infinity or not is obviously mass dependent, e.g.~$\beta=30$ and masses below $0.47$ leads to an infinitely extended cloud. These solutions are unphysical as they give rise to an infinite chemical potential and charge density on the conformal boundary. In contrast, in a physically relevant setting the charge density and chemical potential remains finite. 
The expansion of the cloud when increasing $\beta$, or lowering $m$, should not be interpreted only as the cloud stretching out farther, but also as the scale $T/\mu$ decreasing. If the chemical potential is to remain fixed, the charge and mass of the black hole would decrease when the cloud expands, as the black hole dissipates into the cloud.
The zero temperature electron star limit, $T=0$, is however not attainable in our framework, as we fix the position of the horizon to $z=1$. In this framework, instead the relevant scale $T/\mu$ goes to zero, e.g.~when approaching the critical limit with a diverging $\mu$ (from below in $m$).

While solutions beyond the critical limit $\beta_{crit}(m)$ are unphysical in this setting, solutions close to the limit are more interesting (but also more computationally challenging). Within the cloud, the geometry has a Lifshitz-scaling, which gives rise to a higher speed of light in the UV side of the cloud than in the IR. The further the cloud stretches, the larger this difference becomes.

\subsection{First order perturbation}

Varying the action gives the equations of motion for the respective fields. Rather than solving for the most general field configuration, we work in a static background, isotropic in $(x,y)$, and add wavelike-perturbations and study the linear response.\footnote{We use the same conventions as in previous work~\cite{Aronsson:2017dgf,Aronsson:2018yhi,Gran:2018vdn}, and further detail about the parametrization of our fields is found in appendix~\ref{app:details}.}
Making this ansatz for the fields turns each equation into one equation for the background and one equation for the perturbation. Due to Bianchi-identities, some equations simplify to algebraic constraints which can be used to simplify the remaining equations. After this simplification, there are four non-linear differential equations for the background, and seven linear differential equations for the perturbations.
It should also be noted that we are working in the longitudinal sector, where no magnetic flux is induced, so we do not have to take into account the corrections to the electron cloud model in the presence of magnetic fields~\cite{Puletti:2015gwa}. Further details about the equations of motion we solve are in appendix~\ref{app:details}.

Of the seven perturbation equations, only six are present outside of the cloud region, and the boundary conditions of the seventh equation are set by regularity. For the remaining six second order differential equations, half of their boundary conditions are set at the horizon, by requiring no modes to be outgoing. This means that the modes are either pure gauge or satisfying infalling boundary conditions. 

At the boundary we make the standard choice of Dirichlet boundary conditions for the metric perturbations, as well as for the perturbed $A_t$-component, which ensures that there is no dynamical gravitation on the boundary and fixes the chemical potential in the boundary theory. 
Lastly, one makes a specific choice of boundary condition for the $A_x$-perturbation depending on which features, such as QNMs or CMs, e.g.~plasmon modes, one is interested in computing, corresponding to the boundary conditions
\begin{align}
    \delta\!A_x&=0\,,\label{dirichlet_cond}\\
    \omega^2\delta\!A_x+e\;\delta\!A_x'&=0\,,\label{plasmon_cond}
\end{align}
respectively, where prime denotes the normal derivative at the boundary and $e$ is the boundary Maxwell coupling. In what follows we will set $e=1$, which does not affect the qualitative behaviour of the dispersion relations.

Computationally, since the system is linear, it is more convenient to set arbitrary starting values at the horizon, and find a linear combination of the solutions that satisfy the boundary conditions at $z=0$. In addition to four pure gauge modes, one can specify the infalling solutions as one that is `gravitational' on the horizon, having $\left\{\delta\!g_{xx}^*(1),\delta\!A_x^*(1)\right\}=\{1,0\}$ and one that is `electromagnetic', having $\left\{\delta\!g_{xx}^*(1),\delta\!A_x^*(1)\right\}=\{0,1\}$. Stars indicate the lowest order after Fröbenius expansion.

\section{Results}
\label{sec:results}

In this section we compute the longitudinal CMs and QNMs for three different bulk configurations, and illustrate and discuss how various aspects of the bulk affect physics in the boundary field theory.

\subsection{Schwarzschild background}

\begin{figure}[h]
    \centering
    \begin{minipage}[t]{0.45\textwidth}
        \centering
        \includegraphics[width=0.87\linewidth]{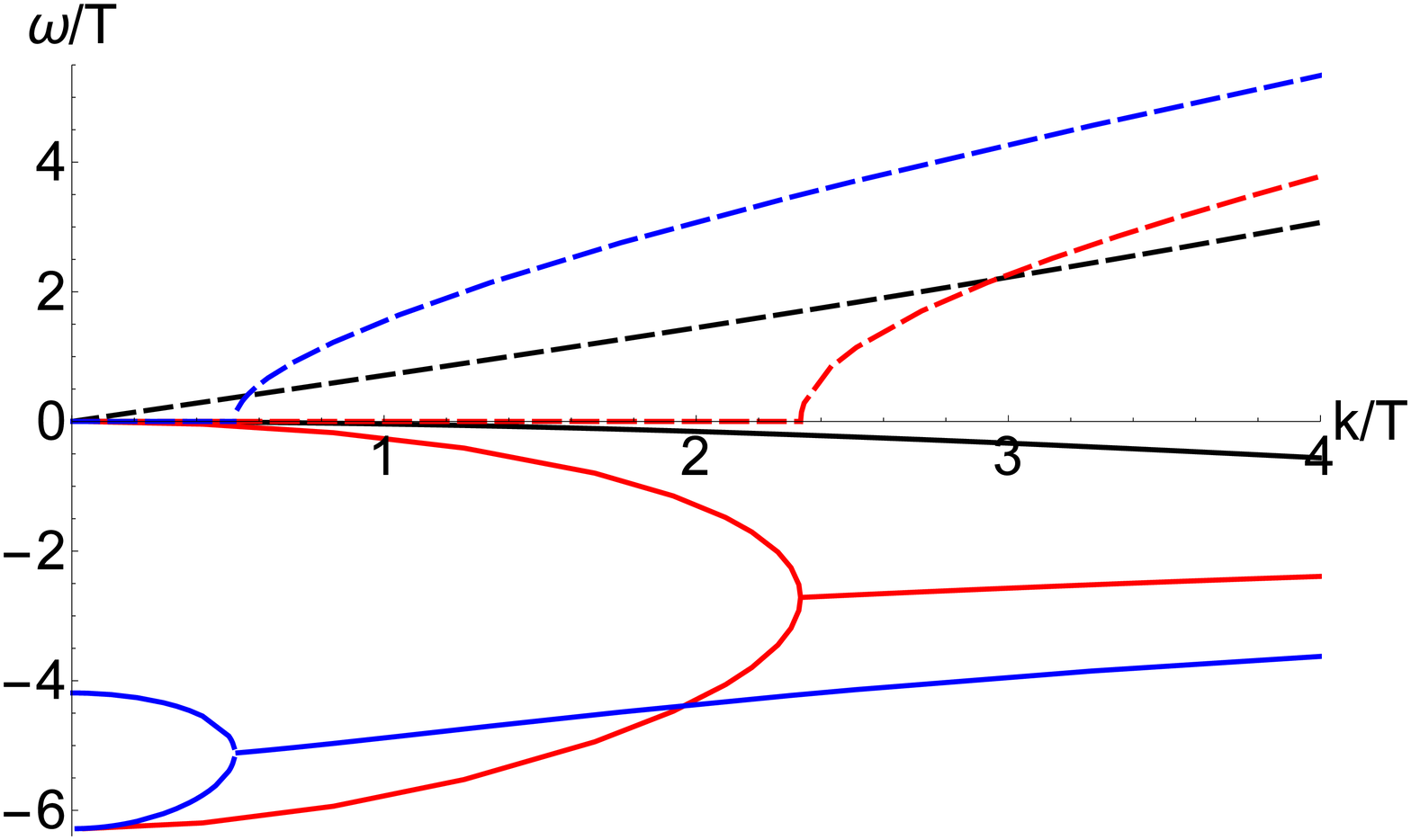}
        \caption{The lowest modes in the Schwarzschild background. The (black) gravitational curve is solely in the gravitational sector and thus both a CM and a QNM. The lowest electromagnetic CM (blue) and QNM (red) are distinct. Real parts are shown as dashed lines and positive to lessen clutter (a negative real part is also a solution). Imaginary parts are solid lines (and always negative).}\label{EC:fig:SchAll}
    \end{minipage}\hfill
    \begin{minipage}[t]{0.45\textwidth}
        \centering
        \includegraphics[width=0.87\linewidth]{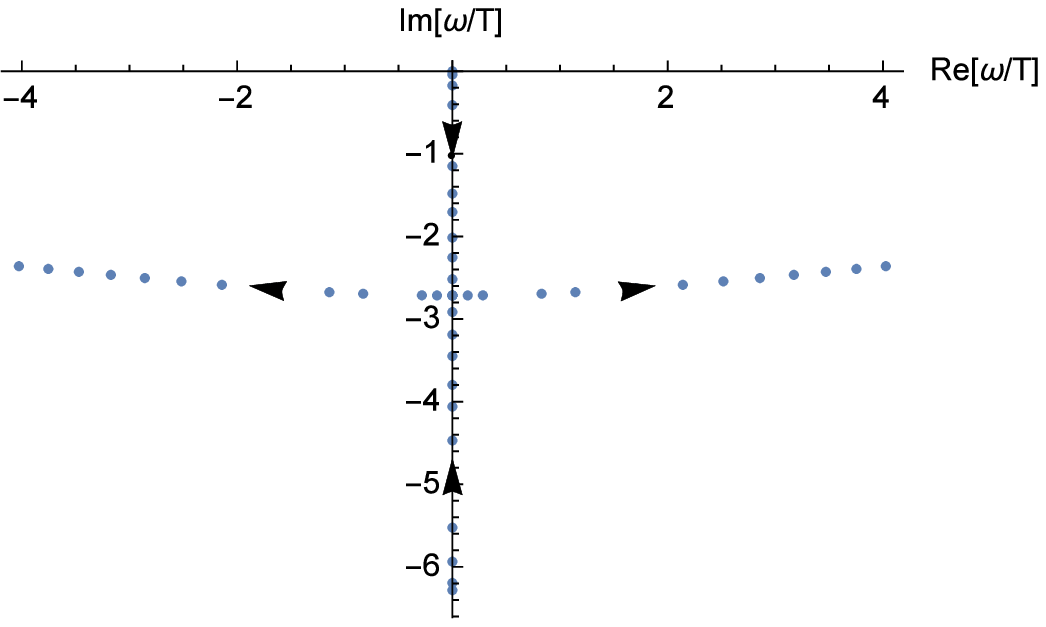}
        \caption{The electromagnetic QNM of \ref{EC:fig:SchAll} but here with $\omega$ in the complex plane, plotted for several different values of $k$. Arrows indicate increasing values of $k$. Note how the two poles moving along the imaginary axis collide and give rise to two symmetric poles about the imaginary axis.}\label{EC:fig:modemerge}
    \end{minipage}
\end{figure}

Setting $Q=0$ means that the chemical potential will be zero everywhere. The absence of background Maxwell-fields will also lead to the decoupling of the 6 perturbation equations into 4 equations only containing the gravitational perturbations and two equations only containing the Maxwell perturbations. This decoupling allows us to identify two different kinds of modes; gravitational and electromagnetic modes.
The lowest gravitational mode (shared for CMs and QNMs) is given in figure \ref{EC:fig:SchAll}, as well as the electromagnetic CMs and QNM. In figure \ref{EC:fig:modemerge} the movement of the QNM poles in the complex $\omega$ plane is displayed.
    
Note that for the CM condition \eqref{plasmon_cond}, only the (gravitational) sound mode exists for small $\omega$ and $k$ whereas for the Dirichlet boundary condition \eqref{dirichlet_cond} there exists a purely diffusive (electromagnetic) mode in addition to the sound mode.

It is also worth noting that whenever the frequency of a mode has a real part, there is a corresponding mode with a negative real part, which is omitted to keep the figures less cluttered. That is, there are technically six modes in e.g.~figure \ref{EC:fig:SchAll}, although for larger $k$ there only appears to be three, as they all have a non-zero real part and hence represent two modes each.

\subsection{Reissner-Nordström background}

\begin{figure}[ht]
    \centering
    \begin{minipage}[t]{0.45\textwidth}
        \centering
        \includegraphics[width=0.87\linewidth]{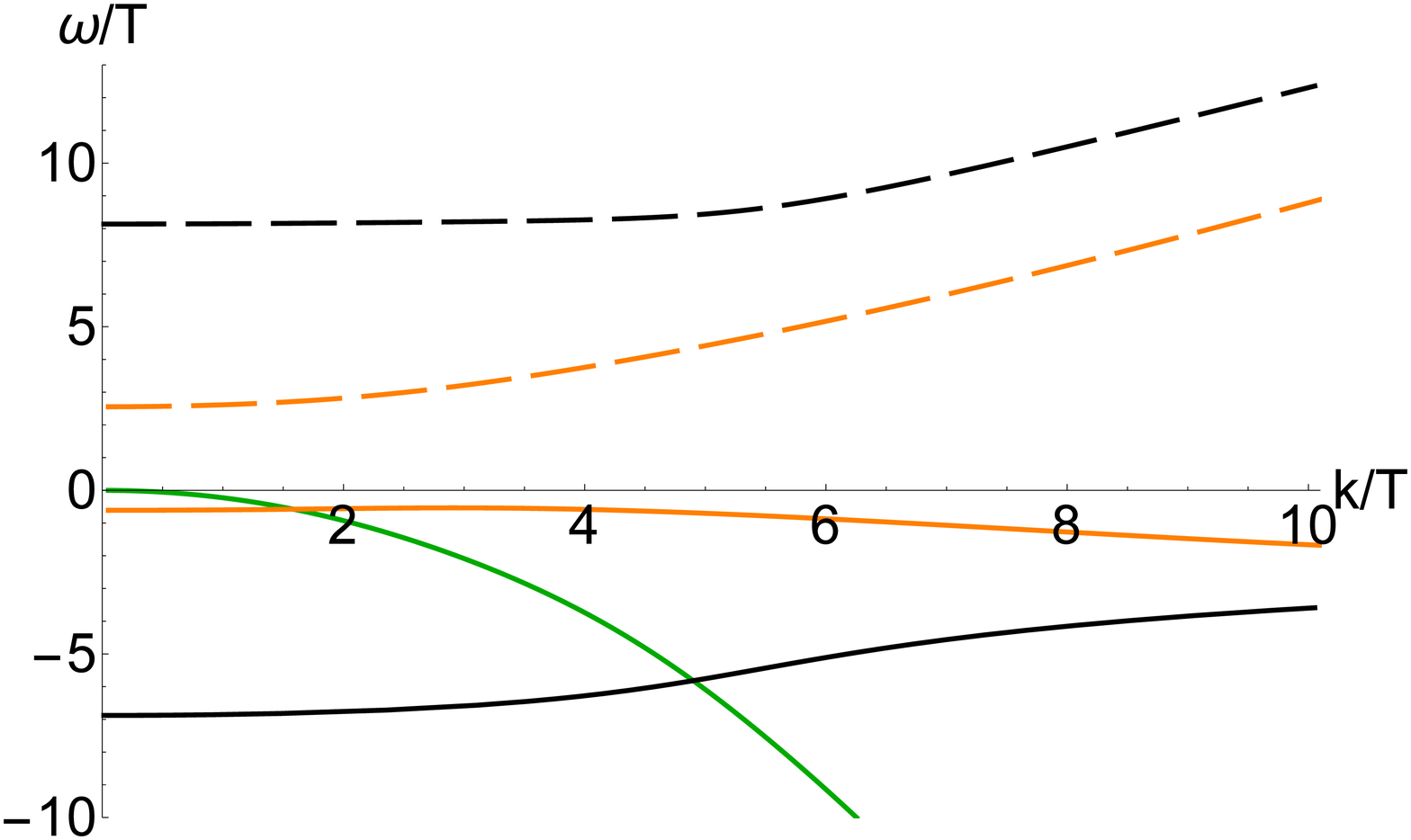}
        \caption{The lowest CM in a $\mu/T\approx5$ RN background. The dashed line is the real part of $\omega/T$, while the solid lines are the imaginary parts. The real part of the green mode coincides with the horizontal axis. The black mode originates from the (gravitational) sound mode, the green from the lowest EM mode.}\label{EC:fig:RNRob}
    \end{minipage}\hfill
    \begin{minipage}[t]{0.45\textwidth}
        \centering
        \includegraphics[width=0.87\linewidth]{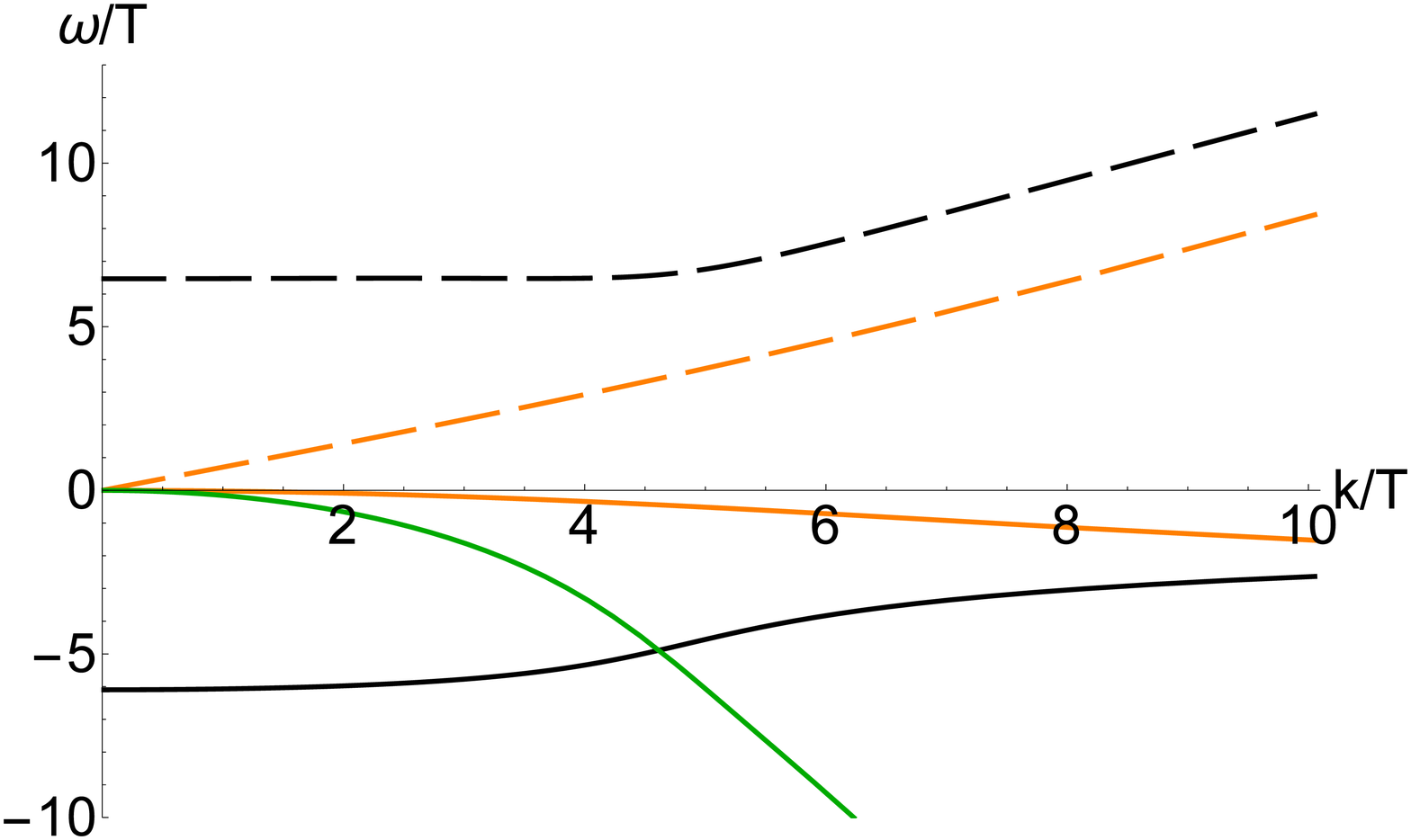}
        \caption{The lowest QNMs in a $\mu/T\approx5$ RN background. The dashed lines are the real parts of $\omega/T$, while the solid lines are the imaginary parts. The real part of the green mode coincides with the horizontal axis.}\label{EC:fig:RNDir}
    \end{minipage}\hfill
\end{figure}

Turning on a non-zero black hole charge Q, and requiring that $m>\mu_{max}(Q)$ or $\beta=0$, leads to the Reissner-Nordström solution. For the specified range of parameters, the chemical potential will not support a charged fluid outside the horizon, but due to the non-vanishing bulk chemical potential the perturbation equations are now coupled. However, as in the Schwarzschild case above, taking the limit $Q\to0$ lets one identify the `origin' of a mode as gravitational or electromagnetic. The lowest CMs and QNMs are given in figures \ref{EC:fig:RNRob} and \ref{EC:fig:RNDir}, respectively. For certain values of the parameters an `exotic' dispersion can be observed\footnote{A similar type of dispersion has previously been observed in the context of QNMs for probe branes in a magnetic field \cite{Jokela:2012vn} (which couples the longitudinal and transverse sectors), for transverse QNMs \cite{Alberte:2017cch}, and recently for longitudinal QNMs in the presence of spacetime filling branes \cite{Gushterov:2018spg}.}, c.f.~\cite{Gran:2018vdn}, which is a leading order effect for the CMs, but subleading for the QNMs due to the presence of the dominating sound mode.

\subsection{Electron Cloud background}

A non-zero black hole charge Q, and a particle mass $m<\mu_{max}$ together with $\beta\neq0$ leads to the `electron cloud' background, where a charged fluid is supported for some radial coordinate range outside the horizon. The numerical analysis gets more complicated in the cloud, and the Lifshitz background allows for a modification of the dispersion relations' slopes. In the limit $\beta\to0$ one gets back the RN-solution, and can thus again trace back the origin of the modes as being gravitational or electromagnetic\footnote{Up to different types of mode merging, thus yielding modes having multiple origins.}. 

We focus our attention on the five lowest modes. However, due to parity, to any mode with a non-zero real part of $\omega$ there also exists a mode with the same real part, but the opposite sign. This means that in practice, we generally study three different modes. 

\begin{figure}[ht]
    \centering
    \begin{minipage}[t]{0.45\textwidth}
        \centering
        \includegraphics[width=1.0\linewidth]{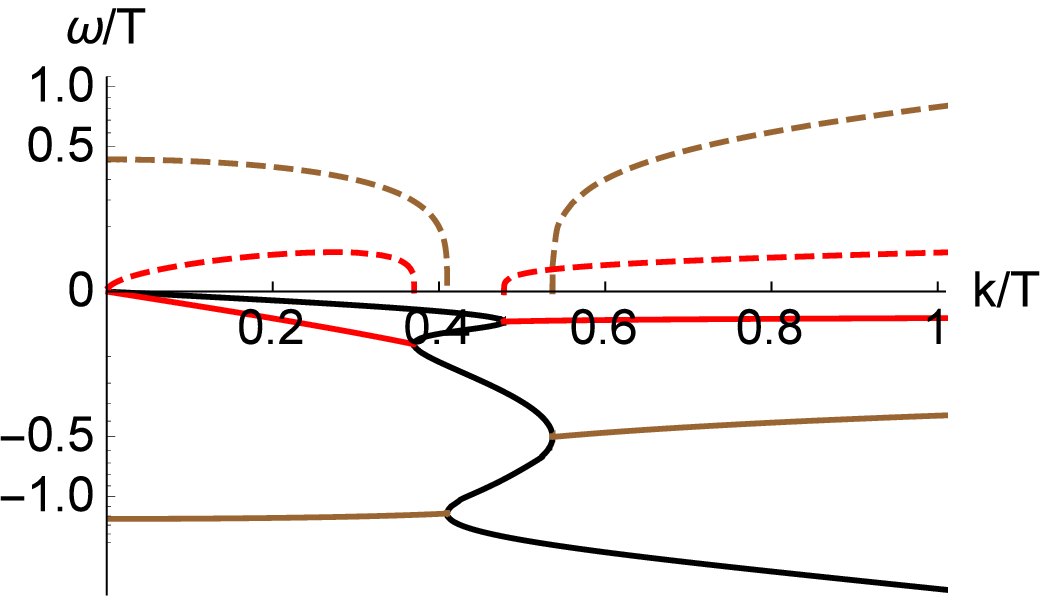}
        \caption{The lowest CMs at $\mu/T\approx1.9$, $m=0.1$ and $\beta=4\pi$. Dashed lines are real parts and solid lines are imaginary parts. Similarly coloured real parts and imaginary parts constitute the same mode. The black line is purely diffusive (and technically five different modes). Note that the $\omega/T$-axis is scaled quadratically to better display all modes, but also to showcase the quadratic imaginary part of the linear mode. }\label{EC:fig:exoticer_cloud}
    \end{minipage}\hfill
    \begin{minipage}[t]{0.45\textwidth}
        \includegraphics[width=1.0\linewidth]{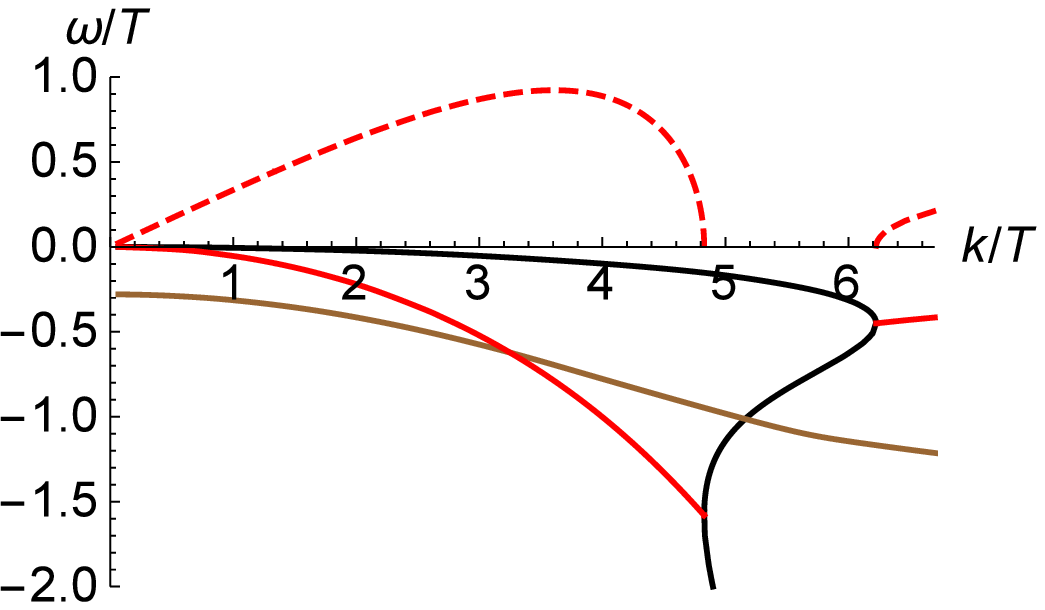}
        \caption{The lowest CMs at $\mu/T\approx14.4$, $m=0.2$ and $\beta=10$. Dashed lines are real parts and solid lines are imaginary parts. Similarly coloured real parts and imaginary parts constitute the same mode. The black line is purely diffusive (and technically three different modes). The brown line is the imaginary part of the lowest gapped mode, whereof the real part has been omitted, as it is an order of magnitude beyond the scale of the other modes.}\label{EC:fig:exotic_cloud}
    \end{minipage}
\end{figure}
The most notable difference compared to the Reissner--Nordström model is the appearance of a new CMs. This mode starts off linearly, meaning that the five lowest modes are one diffusive, two linear and two gapped, much like the case for QNMs. This mode however, exhibits a similar `exotic' behaviour as the gapped CM and QNM, as discussed in~\cite{Gran:2018vdn}, in contrast to the linear QNM, i.e.~the sound mode. The mode with the exotic behaviour can be seen in figure \ref{EC:fig:exotic_cloud}. As for the RN model, this is a leading order phenomenon for CMs, but subleading for QNMs.

The exotic behaviour of all modes can to some extent be manipulated with all three parameters. This ultimately means that for some choices, the exotic behaviours coincide, resulting in very peculiar dispersion relations. One such case is shown in figure \ref{EC:fig:exoticer_cloud}.

Tweaking the parameters appears to have significantly larger impact on the CMs compared to the QNMs. In the following, we highlight some of these differences.

\subsubsection{Varying $\beta$}

\begin{figure}[ht!]
    \centering
        \includegraphics[width=0.44\linewidth]{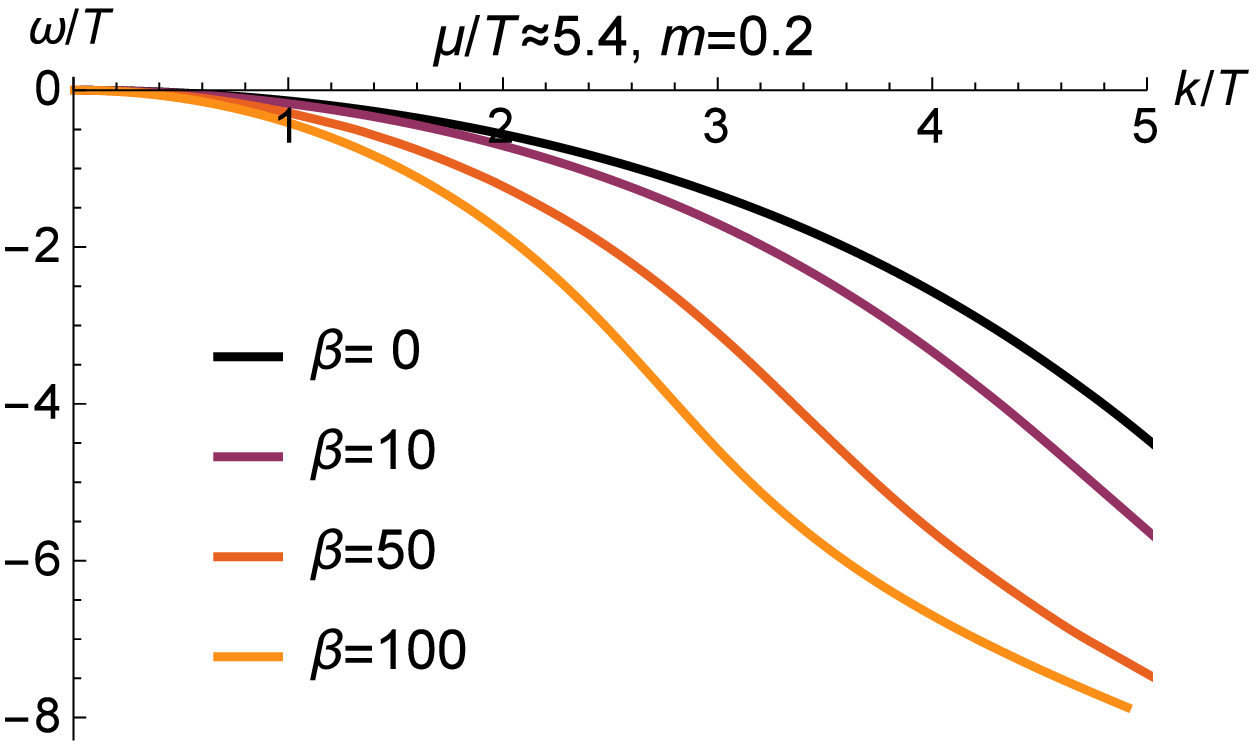}
        \hspace{.05\linewidth}
        \includegraphics[width=0.44\linewidth]{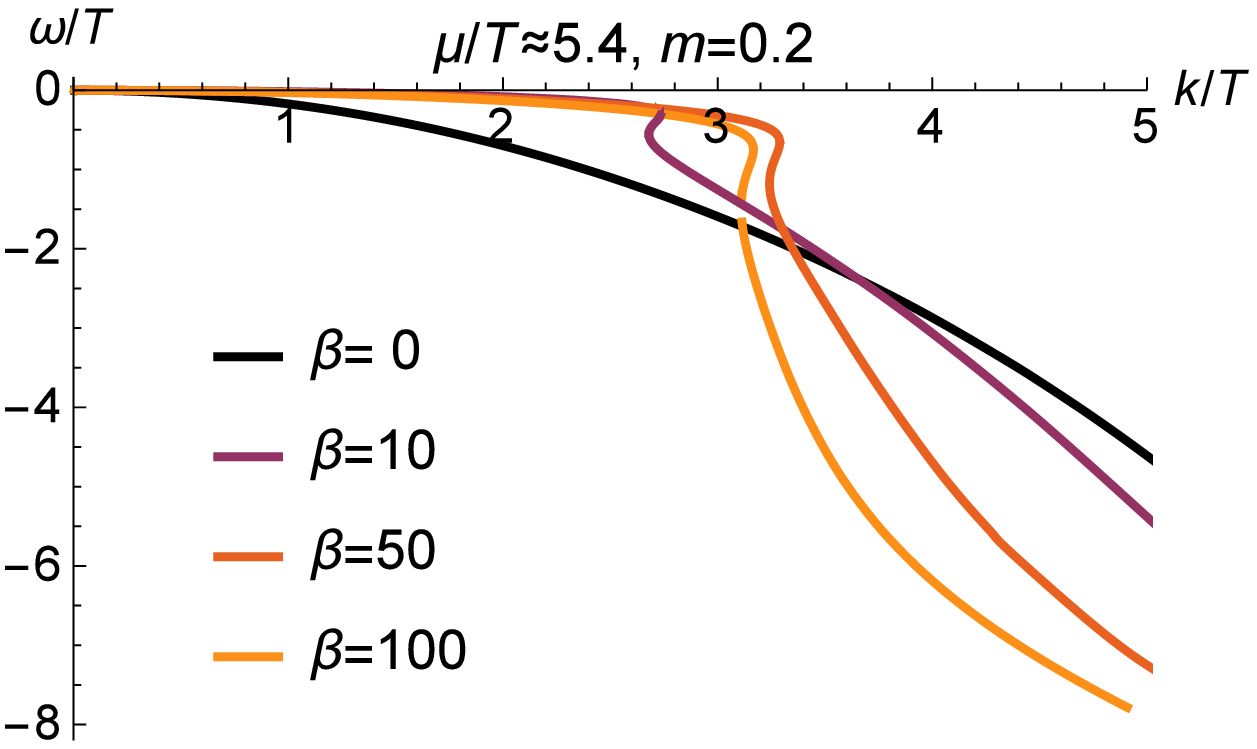}
        \caption{The diffusion QNM (left) and CM (right) at $\mu/T\approx5.4$, $m=0.2$.
        Note the formation of the crease in the CM, which becomes more pronounced as $\beta$ increases.
        }\label{fig:betachange_diff}
\end{figure}

\begin{figure}[ht!]
    \centering
        \includegraphics[width=0.44\linewidth]{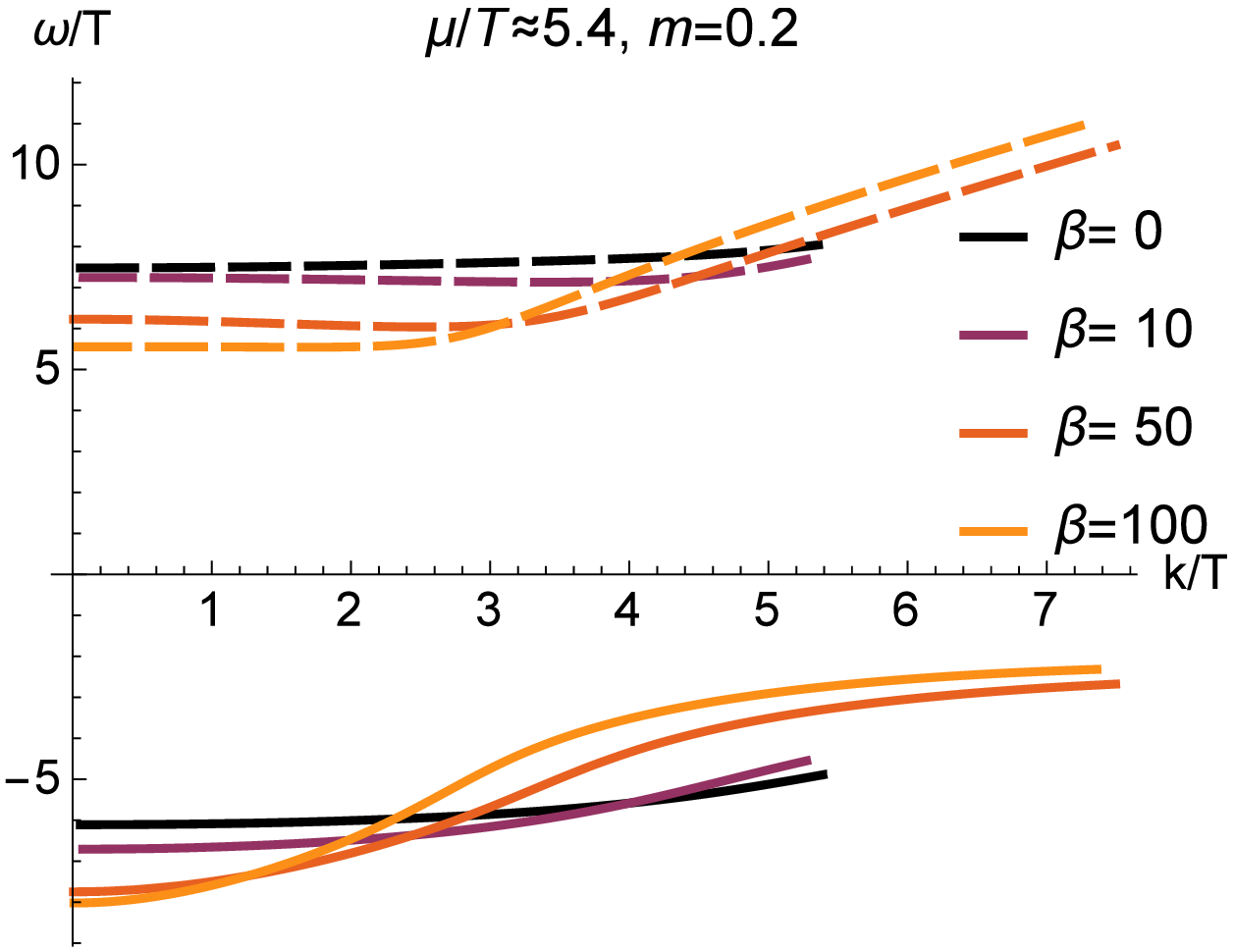}
        \hspace{.05\linewidth}
        \includegraphics[width=0.44\linewidth]{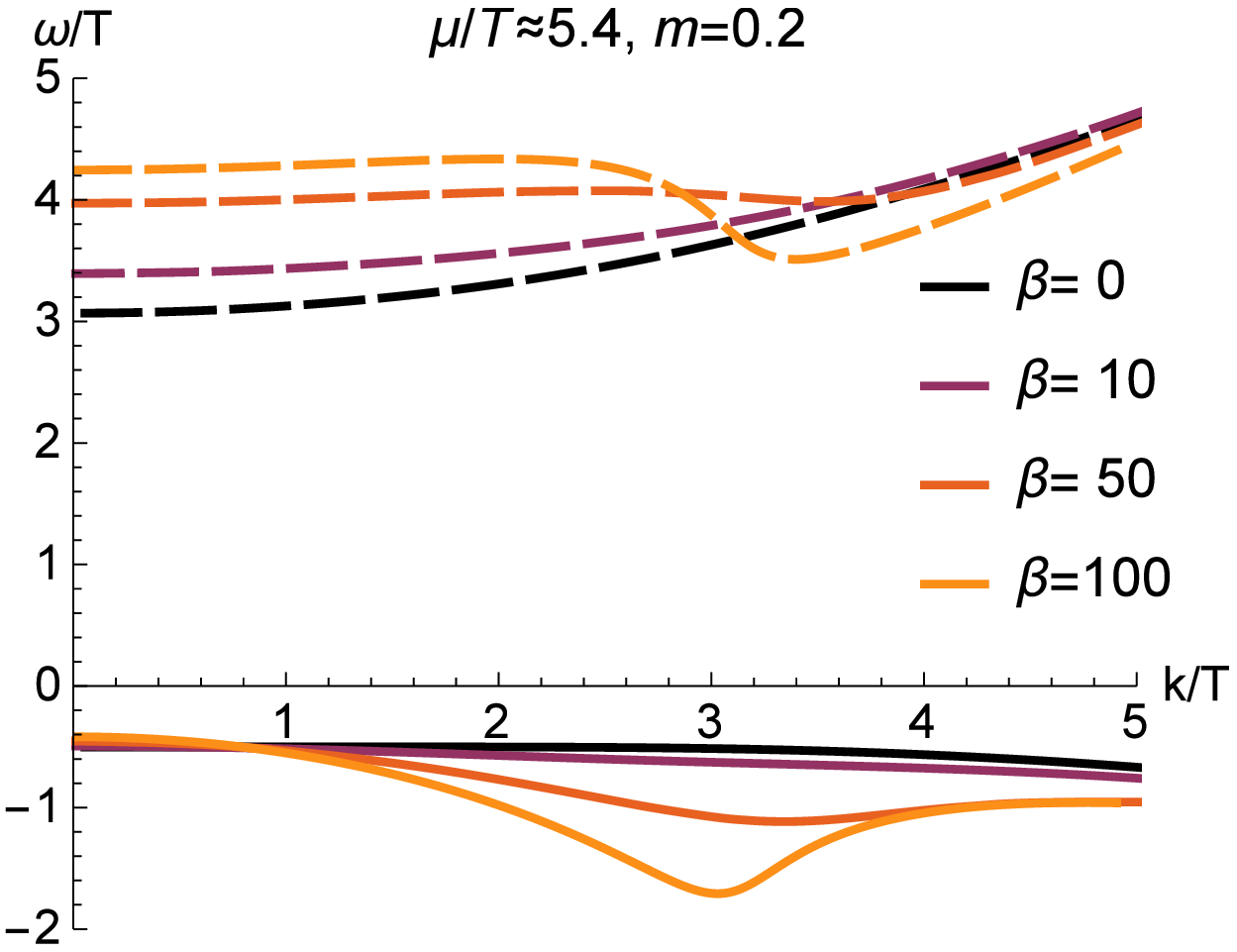}
        \caption{The gapped QNM (left) and CM (right) at $\mu/T\approx5.4$, $m=0.2$.
        }\label{fig:betachange_gap}
\end{figure}

\begin{figure}[ht!]
    \centering
        \includegraphics[width=0.42\linewidth]{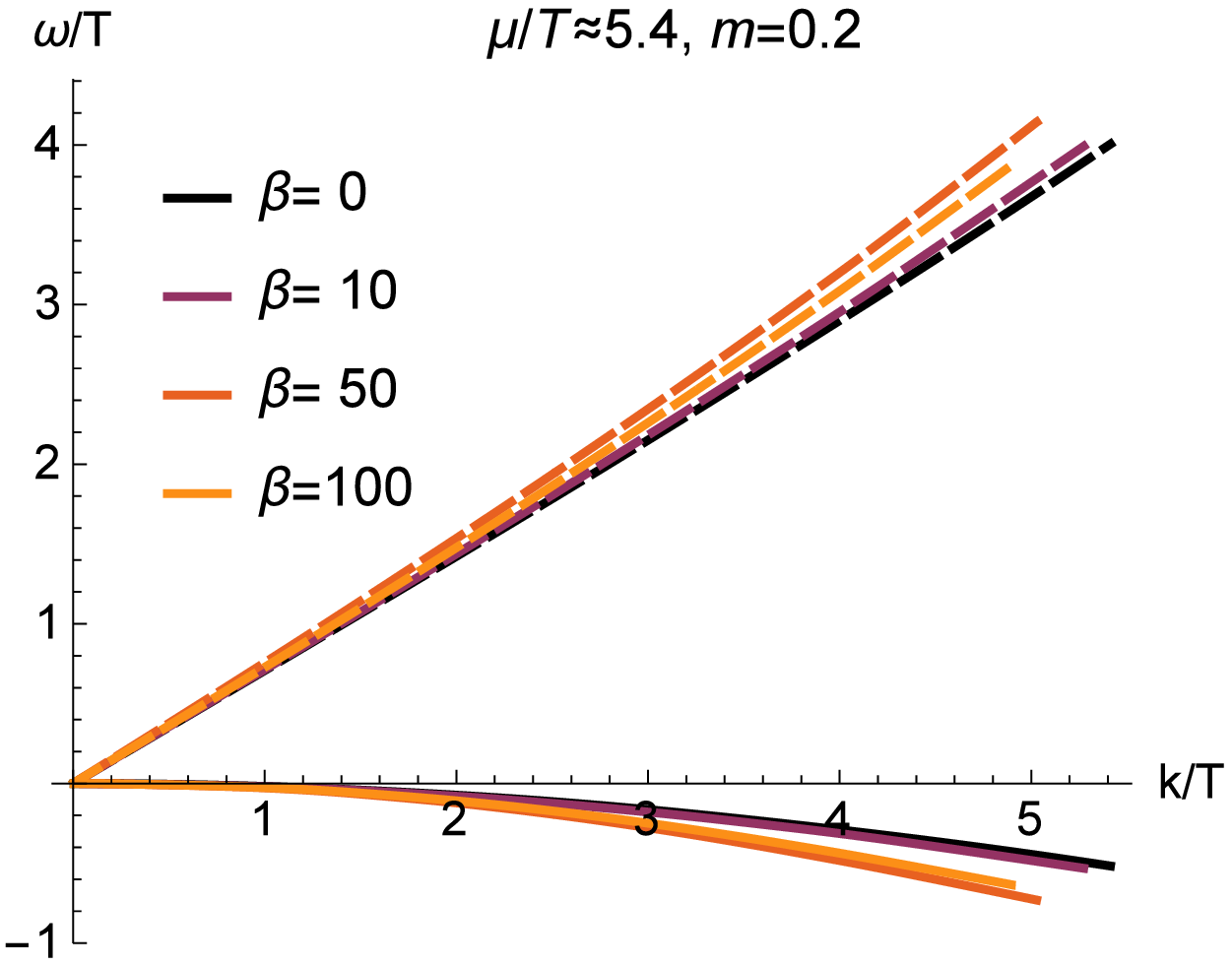}
        \hspace{.05\linewidth}
        \includegraphics[width=0.42\linewidth]{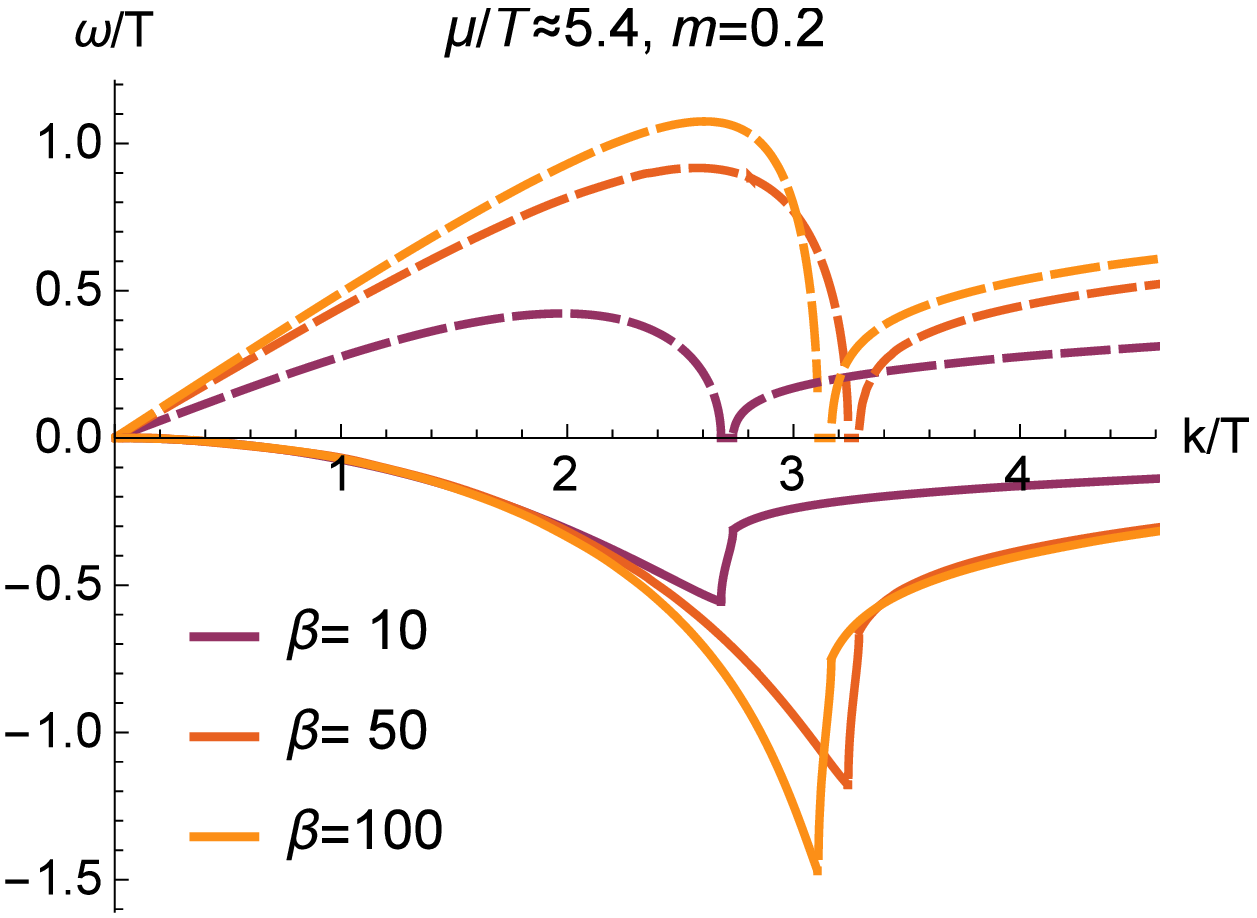}
        \caption{The linear QNM (left) and `cloud' CM (right) at $\mu/T\approx5.4$, $m=0.2$.
        Note the formation of an overdamped `exotic' region in the CM. 
        }\label{fig:betachange_lin}
\end{figure}

Figures \ref{fig:betachange_diff} - \ref{fig:betachange_lin} show the impact of changing $\beta$ without changing the other parameters. Especially noteworthy is the crease in the gapped dispersion, figure \ref{fig:betachange_gap}, indicating that the system is going toward having an exotic gapped mode. Also worth noting is the movement of the exotic region of the linear mode in figure \ref{fig:betachange_lin} at increasing $\beta$, which appears to move towards higher $k/T$ for small $\beta$ but then turns again and moves toward smaller $k/T$ for sufficiently large $\beta$ ($\beta\gtrsim80$).%

\newpage
\subsubsection{Varying $m$}

\begin{figure}[ht!]
    \centering
        \includegraphics[width=0.44\linewidth]{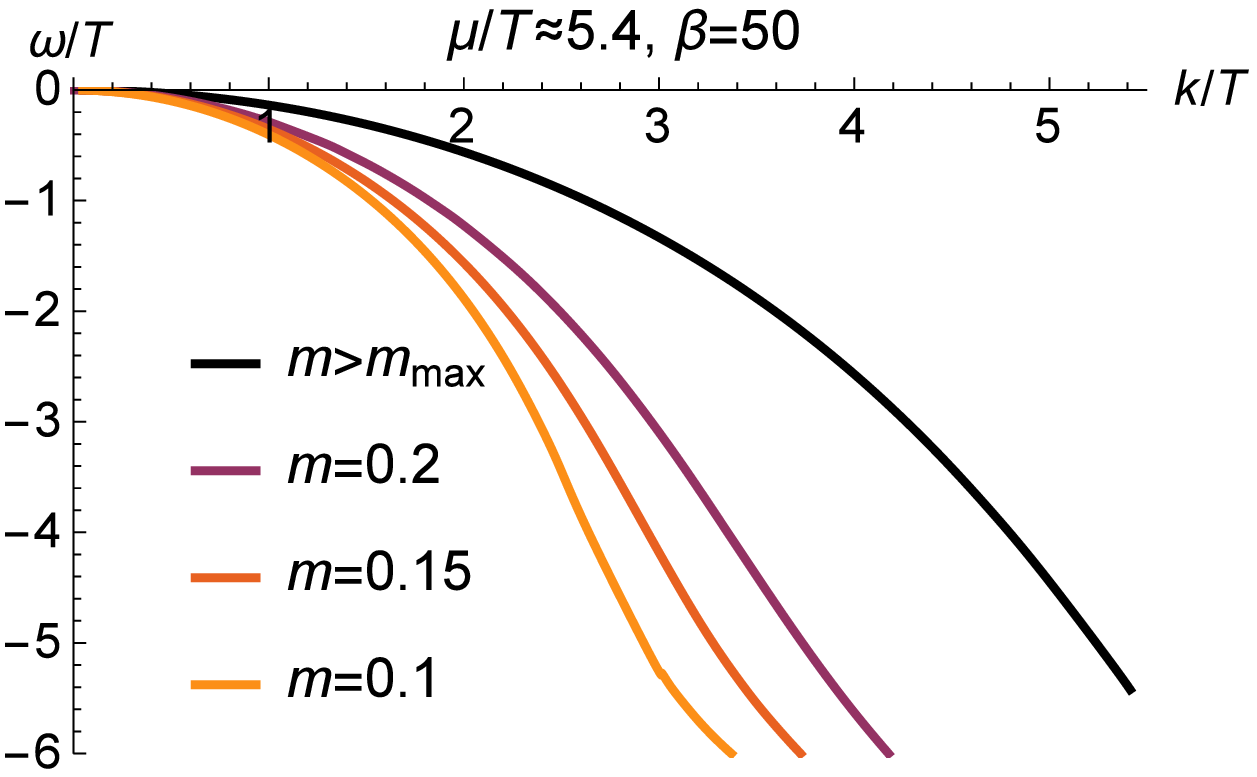}
        \hspace{.05\linewidth}
        \includegraphics[width=0.44\linewidth]{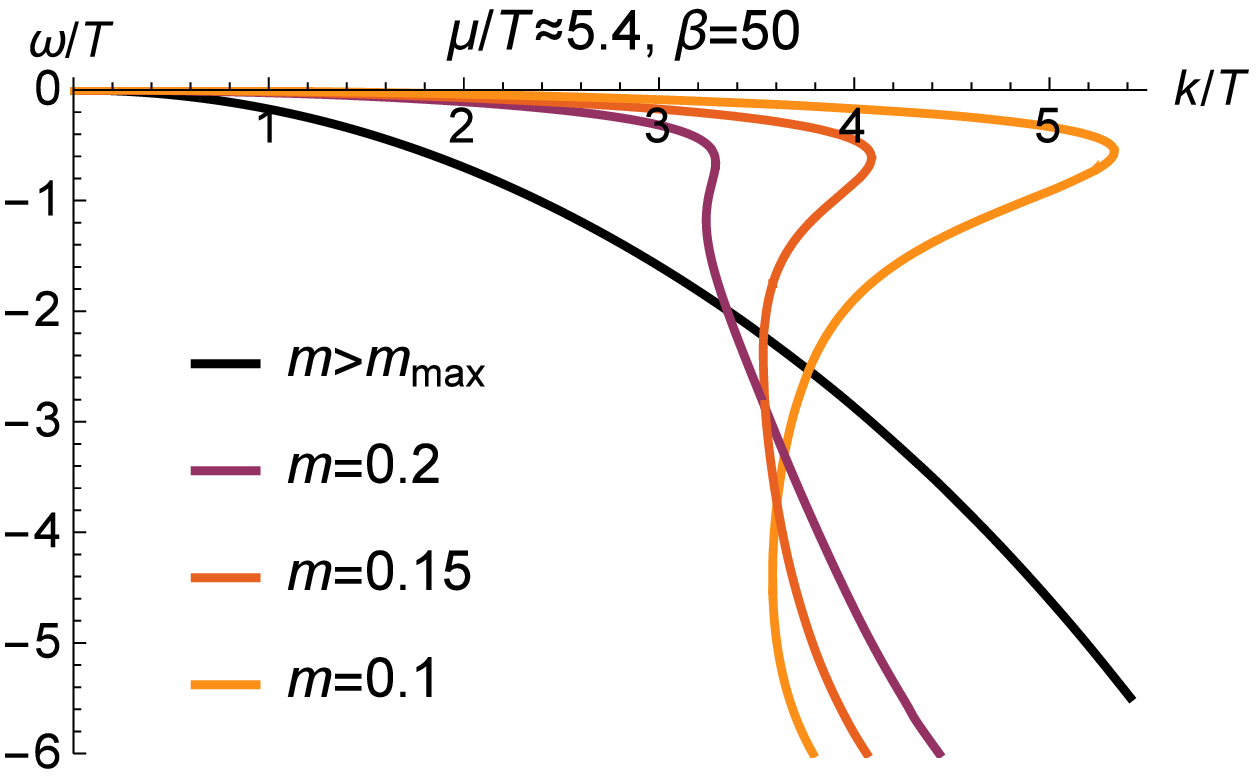}
        \caption{The diffusion QNM (left) and CM (right) at $\mu/T\approx5.4$ and $\beta=50$.
        }\label{fig:mchange_diff}
\end{figure}

\begin{figure}[ht!]
    \centering
        \includegraphics[width=0.44\linewidth]{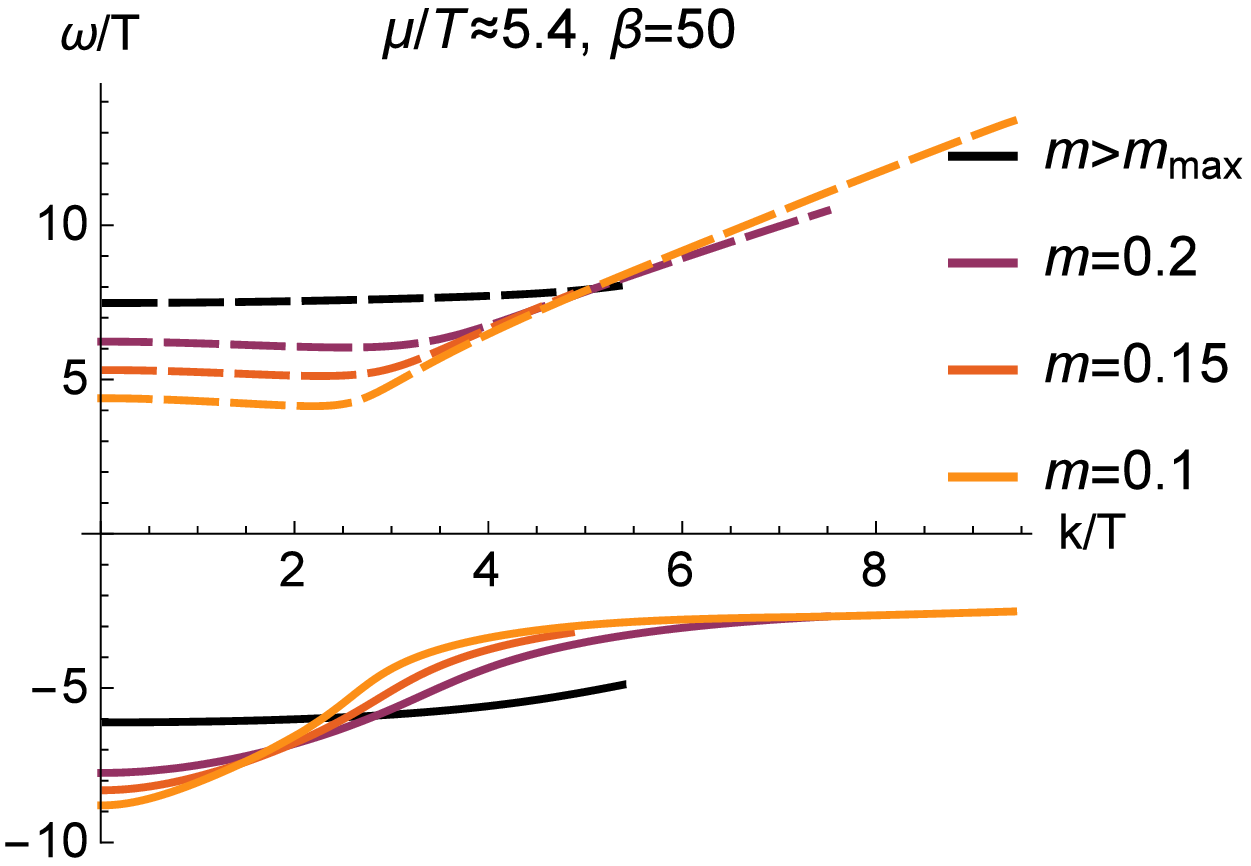}
        \hspace{.05\linewidth}
        \includegraphics[width=0.44\linewidth]{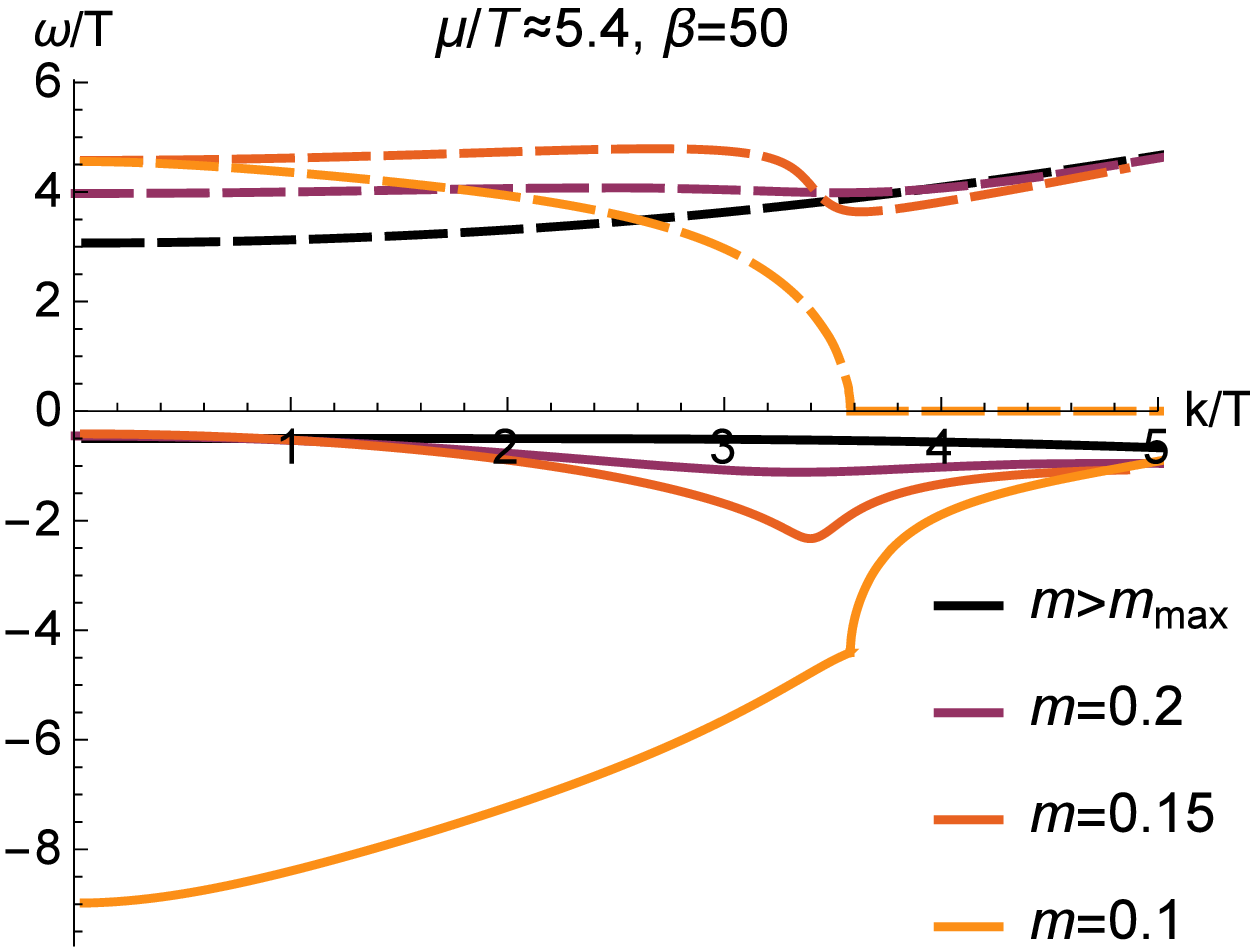}
        \caption{The gapped QNM (left) and CM (right) at $\mu/T\approx5.4$ and $\beta=50$.
        }\label{fig:mchange_gap}
\vspace{-33mm}
\end{figure}

\begin{figure}[ht!]
    \centering
        \includegraphics[width=0.44\linewidth]{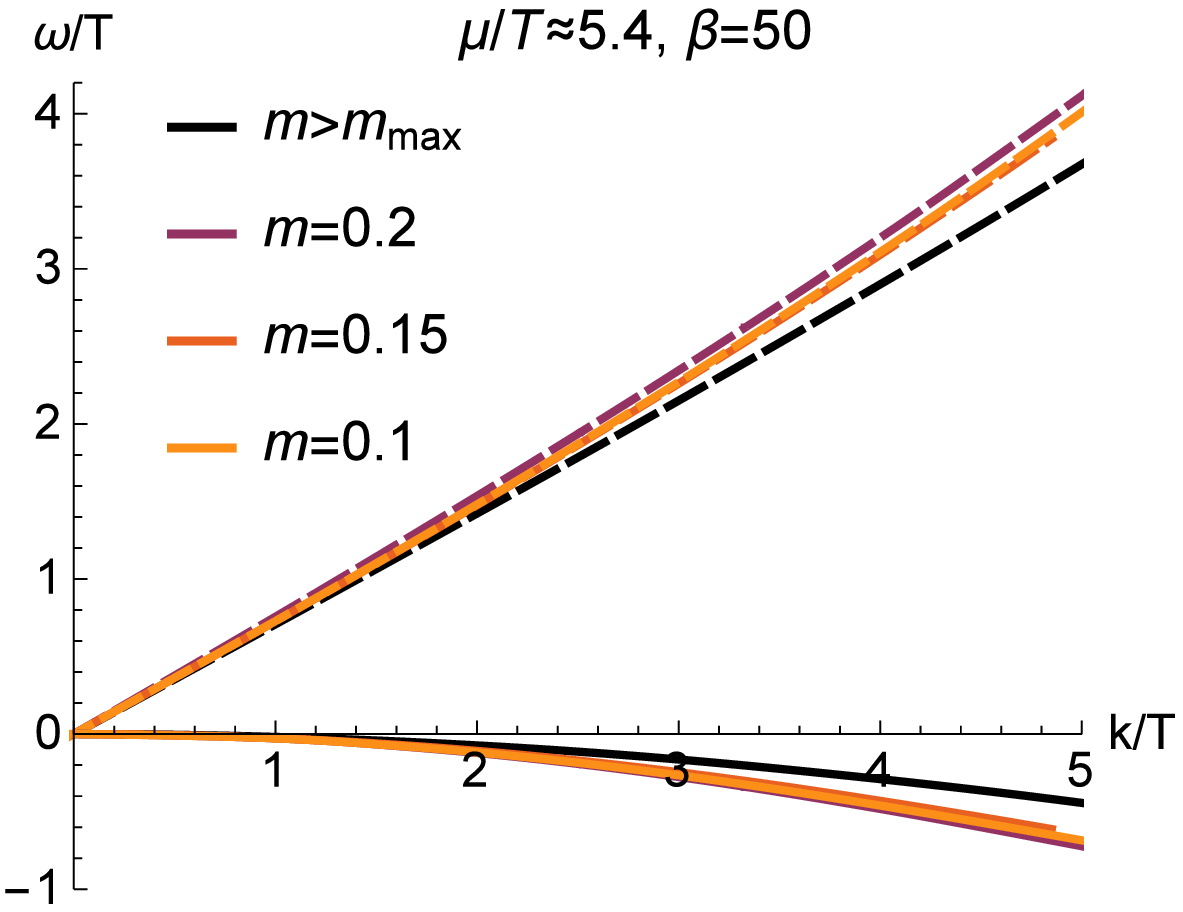}
        \hspace{.05\linewidth}
        \includegraphics[width=0.44\linewidth]{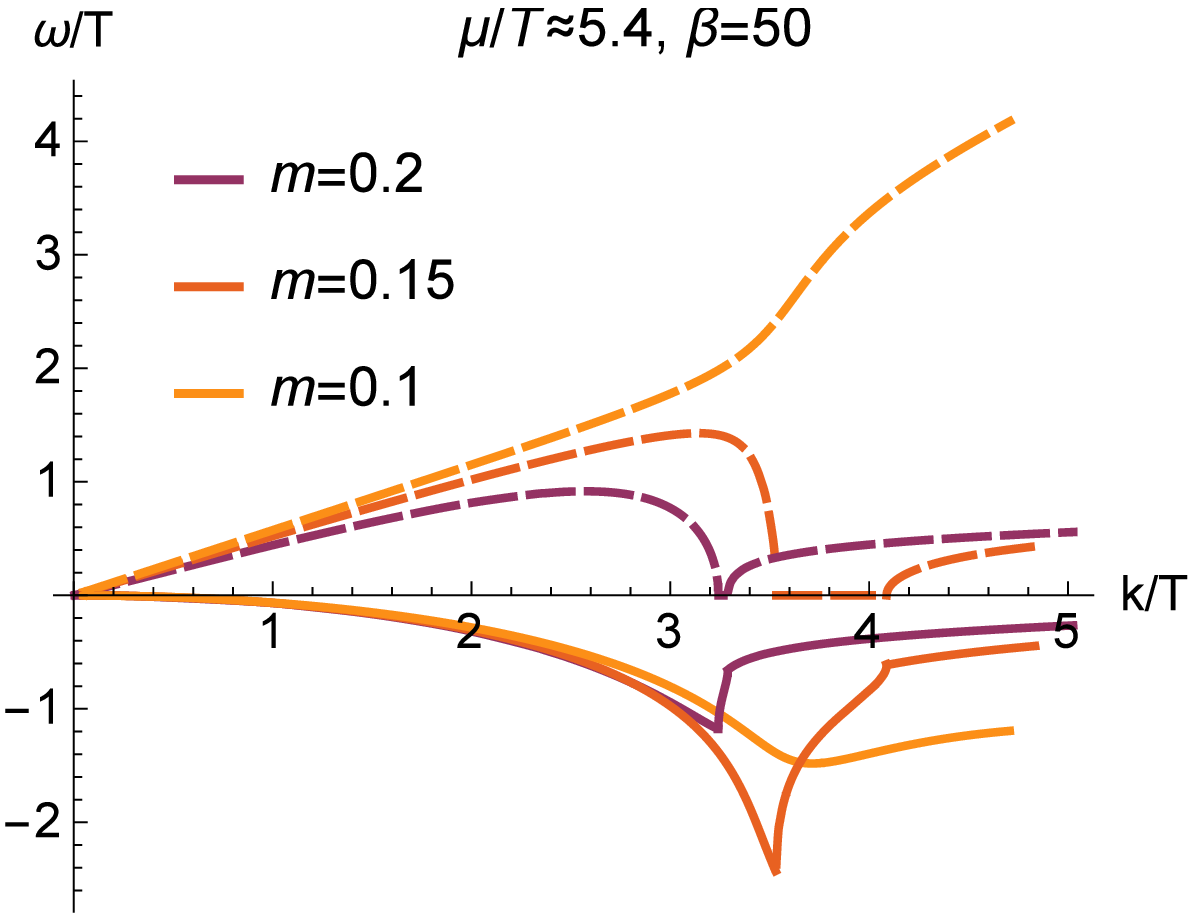}
        \caption{The linear QNM (left) and `cloud' CM (right) at $\mu/T\approx5.4$ and $\beta=50$.
        }\label{fig:mchange_lin}
\end{figure}

Figures \ref{fig:mchange_diff} - \ref{fig:mchange_lin} show the impact of changing $m$ independently of the other parameters. Note especially the appearance of an exotic region for both the gapped mode and the linear mode, and how one is becoming more pronounced by decreasing the mass, while the other is weakened.%

\newpage 

\subsubsection{Varying $\mu/T$}

\begin{figure}[ht!]
    \centering
        \includegraphics[width=0.44\linewidth]{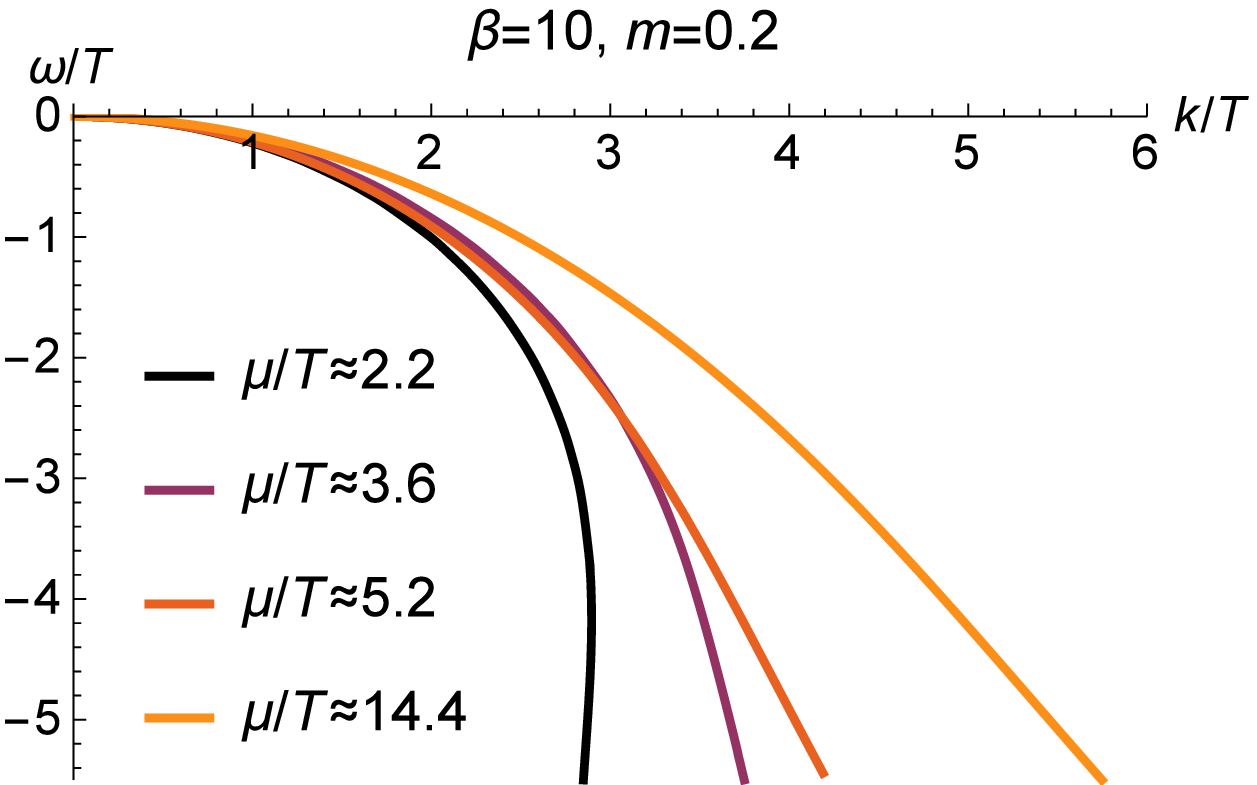}
        \hspace{.05\linewidth}
        \includegraphics[width=0.44\linewidth]{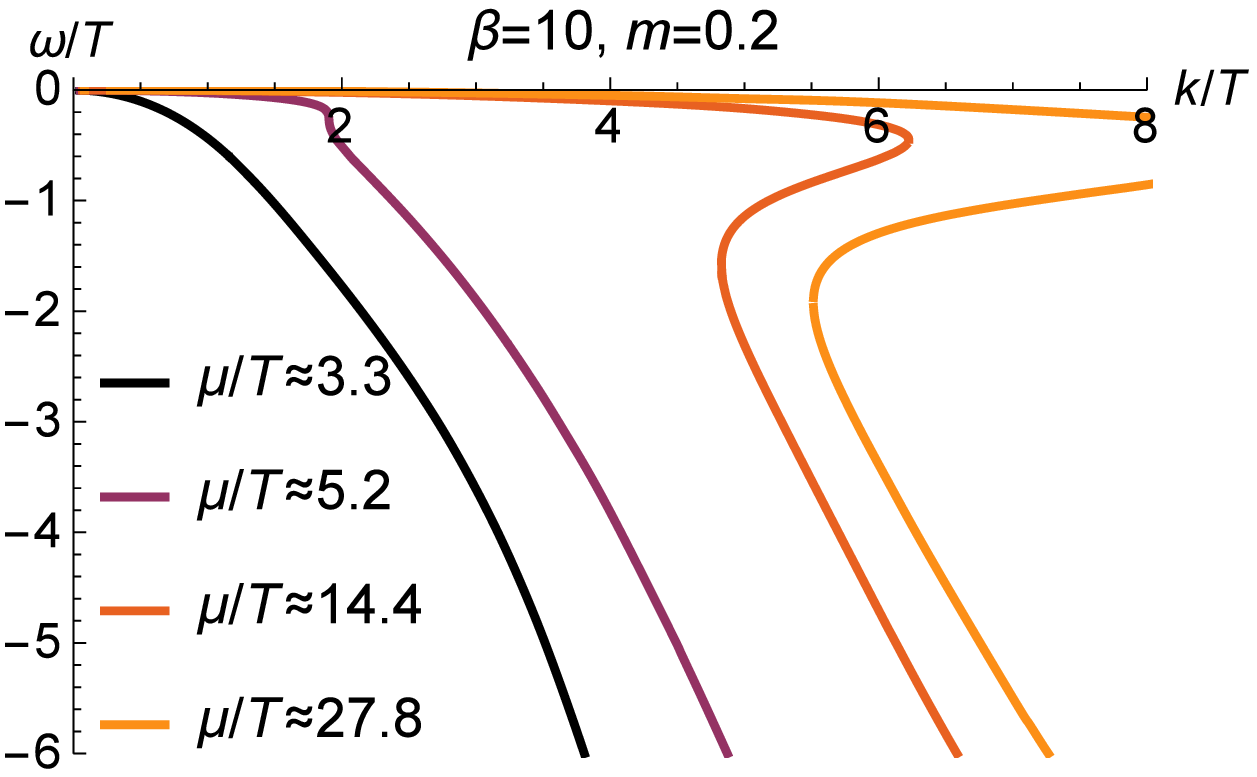}
        \caption{The diffusion QNM (left) and CM (right) at $\beta=10$ and $m=0.2$.
        }\label{fig:Tchange_diff}
\end{figure}

\begin{figure}[ht!]
    \centering
        \includegraphics[width=0.44\linewidth]{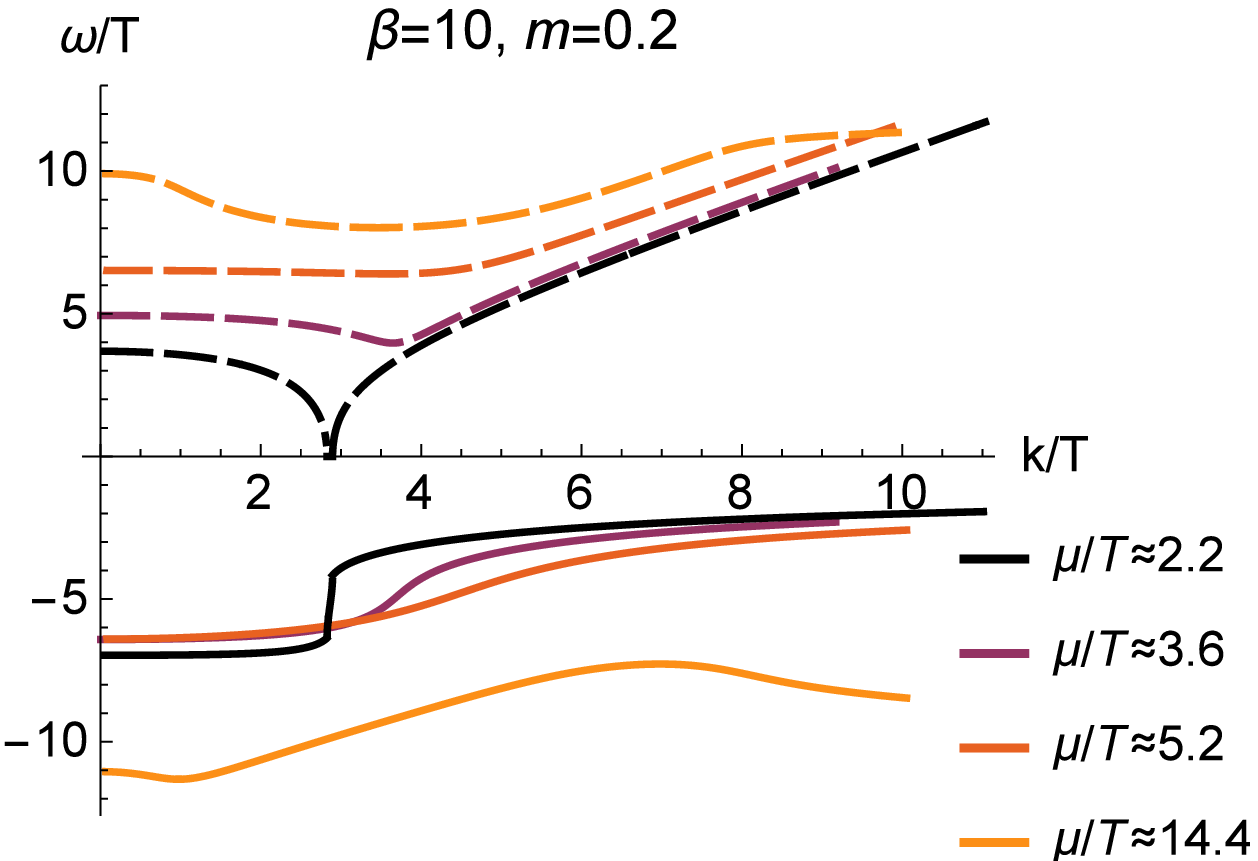}
        \hspace{.05\linewidth}
        \includegraphics[width=0.44\linewidth]{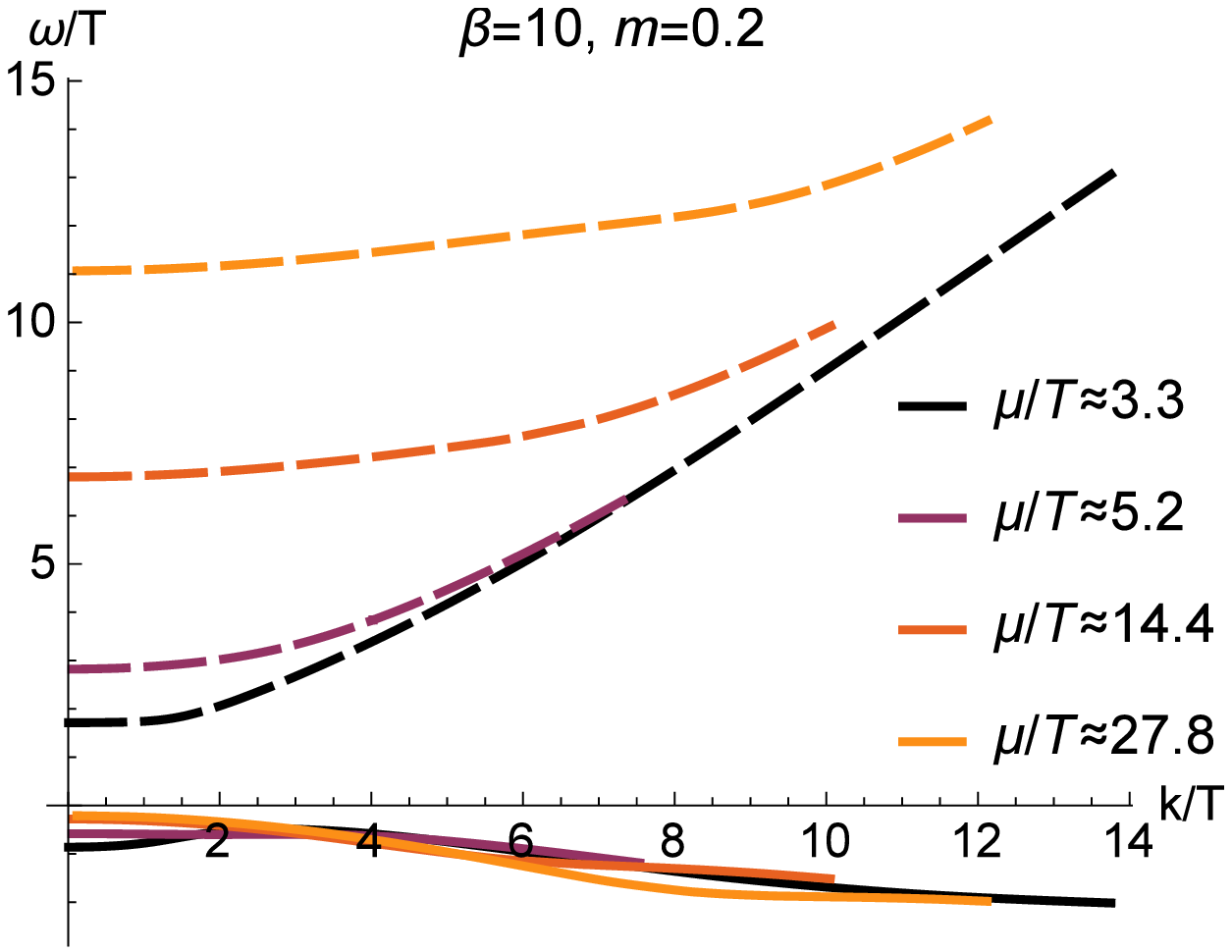}
        \caption{The gapped QNM (left) and CM (right) at $\beta=10$ and $m=0.2$.
        }\label{fig:Tchange_gap}
\end{figure}

\begin{figure}[ht!]
    \centering
        \includegraphics[width=0.44\linewidth]{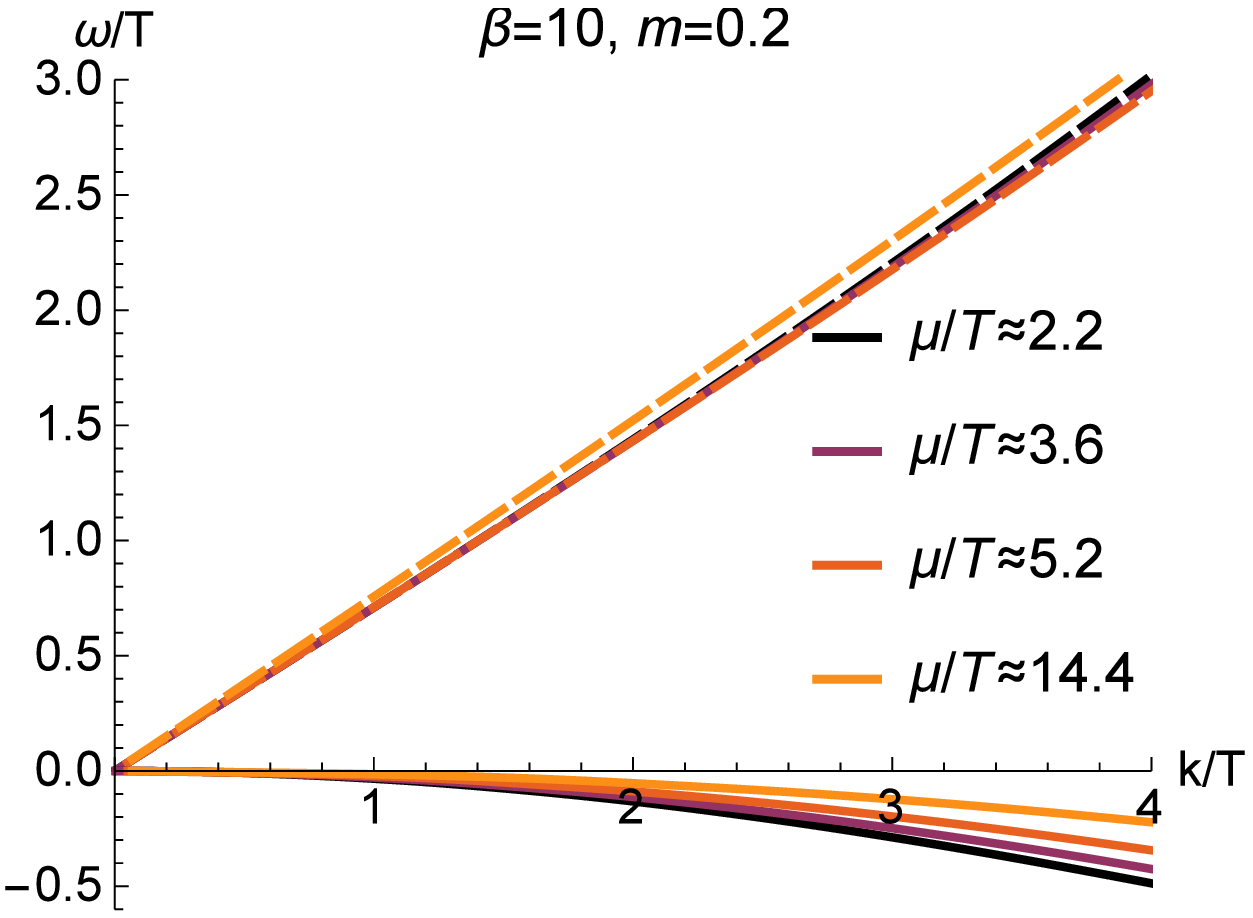}
        \hspace{.05\linewidth}
        \includegraphics[width=0.44\linewidth]{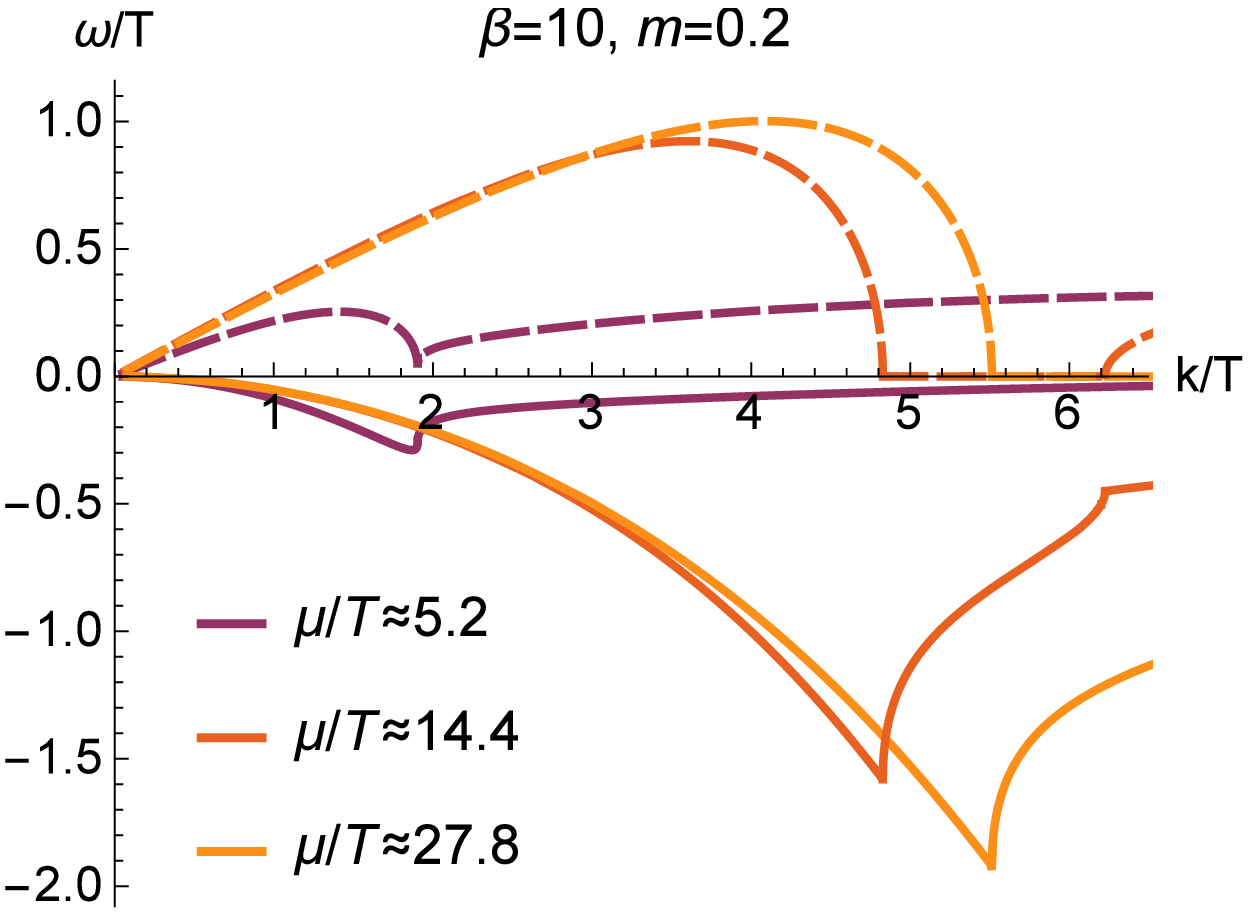}
        \caption{The linear QNM (left) and `cloud' CM (right) at $\beta=10$ and $m=0.2$.
        }\label{fig:Tchange_lin}
\end{figure}

Figures \ref{fig:Tchange_diff} - \ref{fig:Tchange_lin} show the impact of changing $\mu/T$ in the system at fixed $m$ and $\beta$. Note especially the movement of the exotic region in the linear dispersion.
\newpage

\section{Discussion}
\label{sec:discussion}

In this paper we have studied longitudinal QNMs and CMs in the electron cloud background. While QNMs correspond to poles in the screened response function $\chi_{sc}$, the CMs correspond to poles in the physical response function $\chi$ \cite{nozieres1966theory}. The QNMs correspond to Dirichlet conditions for the fields at the asymptotic AdS boundary, while the CMs correspond to a specific type of mixed boundary condition for the Maxwell potential \cite{Aronsson:2017dgf,Aronsson:2018yhi}, related to an RPA form of the Green's function \cite{Witten:2001ua,Mueck:2002gm,Zaanen:2015oix}.

We have previously studied the same type of modes in the Reissner--Nordström background, and found an exotic, or anomalous, dispersion relation which for the CMs is a leading order phenomenon (in the QNM sector there is always a dominating sound mode) \cite{Gran:2018vdn}. More specifically, for certain parameter values of the model there opens up a range of momenta for which the leading propagating mode becomes non-propagating.
However, it is well-known that the RN black hole is unstable at low temperature corresponding to the extremal RN black hole \cite{Lucietti:2012xr,Aretakis:2011ha,Aretakis:2011hc}. One mode of instability involving fermions is towards the formation of an {\em electron star} \cite{Hartnoll:2009ns} for zero temperature, or an {\em electron cloud} \cite{Puletti:2010de,Hartnoll:2010gu} for non-zero temperatures\footnote{Another important mode of instability instead involves bosons and leads, through the condensation of a charged scalar field, to superconductivity \cite{Hartnoll:2008vx,Hartnoll:2008kx}.}. A natural question to ask is what kind of dispersion relation a more realistic model of a holographic metal, like the electron star or cloud models, has, and in particular if the exotic dispersion is an artifact of the instability of the RN model. However, as the exotic dispersion is a high temperature phenomenon it is expected to appear also in the electron cloud model, which is indeed what we find.

As we have shown in this paper, not only does the exotic dispersion persist in the electron cloud model, there is an extra CM, corresponding to excitations within the charged cloud, which in itself exhibits an exotic dispersion. Note also that the range of parameters for which the exotic behaviour appears makes this phenomenon very difficult to access using conventional methods. Therefore, in (strange) metals without quasi-particles, the appearance of a range of momenta for which the leading order longitudinal mode of transport is strongly suppressed is a holographic prediction that might be possible to test in the near future \cite{2017arXiv170801929M}.

Looking ahead, studying the CMs in yet more realistic models would be of interest. In addition, studying other phenomena in which the dynamic charge response plays an important role is now possible as we know how to properly extract this information from the physical response function $\chi$.

\acknowledgments

This work is supported by the Swedish Research Council.

\newpage
\appendix
\addtocontents{toc}{\protect\setcounter{tocdepth}{2}}

\section{Detailed description of the model}\label{app:details}

\subsection{Action}
\label{app:action}

As explained in section \ref{sec:model}, we use the standard AdS-Einstein-Maxwell lagrangian, 
\begin{equation}
    \mathcal{L}_{ME}=- \tfrac{1}{4} F_{\mu \nu} F^{\mu \nu} + \tfrac{1}{2} (-2 \Lambda + R[\triangledown ])\,,
\end{equation}
adding the non-rotating, zero-temperature, ideal fluid lagrangian
\begin{equation}
  \mathcal{L}_{fl}=- \rho_{\text{fl}}(n) +n u^{\mu} A_{\mu} + n u^{\mu} \partial_{\mu}\phi +
  \lambda\left(1+ g_{\mu\nu}u^{\mu} u^{\nu}\right)\,,\label{apeq:fl-lag}
\end{equation}
to describe the `cloud'.
Here $\rho_{fl}$ is the energy density and $n$ is the number density, and the charge of the fluid particles has been set to unity. The `Clebsch' potential $\phi$ ensures that the mass of the fluid is conserved. Similarly, $\lambda$ is a Lagrange multiplier implementing that $u^{\mu}$ squares to minus one, as required for the 4-velocity of the fluid.

The fluid Lagrangian can be rewritten as
\begin{equation}
  \mathcal{L}_{fl}=-\rho_{\text{fl}}+n \mu + \lambda\left(1+ g_{\mu\nu}u^{\mu} u^{\nu}\right)
\end{equation}
with the local \emph{bulk} chemical potential $\mu$ defined as
\begin{equation}
  \mu=u^{\mu}\left( A_{\mu}+\partial_{\mu}\phi\right).
\end{equation}

\subsubsection{Equations of motion}
By varying the action with respect to each of the independent fields we get the following equations of motion,
\begin{align}
    \lambda:\; 0=&1+ g_{\mu\nu}u^{\mu} u^{\nu}\label{EC:var:lambda00}\\
    n:\; 0=&-\frac{\partial\rho_{fl}}{\partial n}+\mu\\
    \phi:\; 0=& - \nabla_{\mu}\left(n u^{\mu}\right)\\
    u:\; 0=&         A_{\mu} n + 2 \lambda u_{\mu}  + n \partial_{\mu}\phi \label{EC:var:u}\\
    A:\; 0=&    - n u^{\mu} -  \nabla_{\kappa}\nabla^{\kappa}A^{\mu} + \nabla_{\kappa}\nabla^{\mu}A^{\kappa}\\
    g:\; 0=&    - \Lambda g^{\mu \nu} + \mathcal{L}_{\text{fl}} g^{\mu \nu} -  \tfrac{1}{4} g^{\mu \nu} F_{\kappa \lambda} F^{\kappa \lambda} + F^{\mu \kappa} F^{\nu}{}_{\kappa} -  R[\triangledown ]^{\mu \nu} \nonumber \\ 
 &+ \tfrac{1}{2} g^{\mu \nu} R[\triangledown ] + 2\lambda u^{\mu} u^{\nu}\label{EC:var:g}\,.
\end{align}

These equations can be simplified somewhat. By contracting \eqref{EC:var:u} with $u$ and solving for $\lambda$, using \eqref{EC:var:lambda00}, we get
\begin{equation}
  \lambda=\frac{1}{2}nu^{\mu}\left( A_{\mu}+\partial_{\mu}\phi-s_{\text{pp}} \partial_{\mu}\theta+\alpha \partial_{\mu}\beta\right)=\frac{1}{2}n\mu\,.\label{EC:var:lambda00sol}
\end{equation}
Introducing $p=\mathcal{L}_{\text{fl}}$ and using \eqref{EC:var:lambda00sol},  \eqref{EC:var:g} can be written as
\begin{align}
&&- \Lambda g^{\mu \nu}-  R[\triangledown ]^{\mu \nu} + \tfrac{1}{2} g^{\mu \nu} R[\triangledown ]\nonumber
 -  \tfrac{1}{4} g^{\mu \nu} F_{\kappa \lambda} F^{\kappa \lambda} + F^{\mu \kappa} F^{\nu}{}_{\kappa}  \nonumber \\ 
&& + p g^{\mu \nu} + n\mu u^{\mu} u^{\nu}&=0,
\end{align}
where
\begin{equation}
  p=\mathcal{L}_{\text{fl}}=-\rho_{fl}+\mu n.
\end{equation}

\subsubsection{Ansatz for the fields}
Rather than looking for the most general solution to the equations of motion, we are going to assume a static background, isotropic in $(x,y)$, and add wavelike-perturbations studying the resulting linear response. Making this ansatz turns each equation of motion above into one equation for the background and one equation for the perturbation. 

The background metric is assumed to be a generalization of a planar black hole,
\begin{equation}
    L^{-2}\d s^2=-f(z)\d t^2+z^{-2}\d x^2+z^{-2}\d y^2+g(z)\d z^2,\label{eq:backgroundmetric}
\end{equation}
where the horizon is located at $z=1$ and the AdS boundary at $z=0$, that is $z=r_H/r$ where $r_h$ is the horizon radius and the other coordinates have been rescaled accordingly.

To this we also add perturbations to the metric, invariant under parity $y\to-y$, and work in radial gauge, $\delta g_{z\mu}=0$,

\begin{equation}
  \delta\! g_{\mu\nu}=\varepsilon L^2 e^{-i \omega + k x} \left(\begin{array}{cccc}
\delta\! g_{\text{tt}}(z) & \delta\! g_{\text{tx}}(z) & 0 & 0\\
\delta\! g_{\text{tx}}(z) & \delta\! g_{\text{xx}}(z) & 0 & 0\\
0 & 0 & \delta\! g_{\text{yy}}(z) & 0\\
0 & 0 & 0 & 0
\end{array}\right).
\end{equation}
Similarly, for the Maxwell-potential the static isotropic background, together with radial gauge, leads to

\begin{equation}
    A_{\mu}=L^2\left(\begin{array}{cccc}
    h(z) & 0 & 0 & 0
    \end{array}\right).
\end{equation}
The perturbation then becomes
\begin{equation}
    \delta\! A_{\mu}=\varepsilon L^2 e^{-i \omega + k x}\left(\begin{array}{cccc}
    \delta\! A_t(z) & \delta\! A_x(z) & 0 & 0
    \end{array}\right).
\end{equation}
From \eqref{EC:var:lambda00} and \eqref{eq:backgroundmetric} it is clear that the static solution for $u$ is
\begin{equation}
    u^\mu=L^2\left(\begin{array}{cccc}
    \sqrt{f(z)} & 0 & 0 & 0
    \end{array}\right),
\end{equation}
together with the perturbation
\begin{equation}
    \delta\! u^{\mu}=\varepsilon L^2 e^{-i \omega + k x}\left(\begin{array}{cccc}
    \delta\! u^t(z) & \delta\! u^x(z) & 0 & \delta\! u^z(z)
    \end{array}\right).
\end{equation}
The remaining scalars can be written as
\begin{align}
n&=\sigma(z)+\varepsilon e^{-i \omega + k x} \delta \sigma(z)\, ,\\
\phi&=\phi_0(z)+\varepsilon e^{-i \omega + k x} \delta\! \phi(z)\,,
\end{align}
to first order in the perturbation, and analogously for $\rho_{fl}$
\begin{align}
    \rho_{fl}(n)&=\rho_0(\sigma(z))+\varepsilon e^{-i \omega + k x} \frac{\partial \rho}{\partial n}\delta\sigma(z)\\
    &=\rho_0(z)+\varepsilon e^{-i \omega + k x} \mu_0 \delta\sigma(z)\,.
\end{align}
Inserting the ansatz above into the equations of motion gives us four independent non-trivial background equations,
\begin{align}
    (p_0 + \rho_0) f' - 2 \sqrt{f} \sigma_0 h' + 2 f p'\, ,\\
    4 f g + f g^2 (p_0 + \rho_0) z^2 + g z f' + f z g'\, ,\\
    -2 f + 6 f g z^2 + 2 f g p_0 z^2 + 2 z f' -  z^2 \bigl(h'\bigr)^2\, ,\\
    -2 \sqrt{f} g \sigma + z (p_0+\rho_0) g h' + 2 h''\, ,
\end{align}
from the $g_{tt},\,g_{xx},\,g_{zz}$ and $A_t$ variations.
Furthermore, we get six second order differential equations for the perturbations from the $g_{tt},\,g_{tx},\,g_{xx},\,g_{yy},\,A_t$ and $A_x$ variations, as well as four constraint equations from the $g_{tz},\,g_{xz},\,g_{zz}$ and $A_z$ variations. The constraint equations should be satisfied automatically by any solution of the first six, meaning that they serve only as a check of the solutions. All these equations can be found in appendix \ref{app:pert.eq}.

The expressions resulting from the variations of the other fields can be condensed into simple rules for each of the different perturbations, apart from the $n$-variation which leaves us with a differential equation for $\phi$, effectively getting its derivatives from the expression of $u_z$. This equation can also be found in appendix \ref{app:pert.eq}.

We are thus left with seven independent second order differential equations for the perturbations as well as four background equations we need to solve.

\subsection{Background solution}
\label{app:bgsol}

As explained in the main text, the background will naturally be divided into three regions; a pure RN-solution in the IR, an intermediate cloud-solution and another pure RN-solution in the UV. 
This follows as a RN-solution has a local \emph{bulk} chemical potential that is zero at both the horizon and the boundary, with a maximum in between. When the chemical potential is large enough to support the charged fluid (that is, larger than the mass of the fluid particles), there will be a cloud, located in that region.
Note that the cloud itself affects the region in which a cloud is supported, so one cannot find both the inner and outer bound of the cloud from the pure RN-solution at the center.

\subsubsection{Region I: Inner RN}

The background solution is well known and poses no difficulties, it is the standard Reissner-Nordström solution.
The fluid quantities pressure, number density and energy density, are all zero.

There are four constants determining the Reissner-Nordström solution: the chemical potential $\mu_{\text{RN}}$, the charge $Q$, the mass $M$ and the speed of light $c$. 

\begin{align}
    f(z)&=\frac{c^2}{z^2}-M z+\frac{1}{2} Q^2 z^2\,,\\
    g(z)&=\frac{c^2}{z^4 f(z)}\,,\\
    h(z)&=\mu_{\text{RN}}-Q z\,.
\end{align}
Rescaling the radial coordinate such that the horizon of the black hole is located at $z=1$ leads to the following expression for the mass
\begin{equation}
  f(1)=0\Rightarrow M_1=c_1^2+\frac{1}{2}Q_1^2.
\end{equation}
The chemical potential in the inner region is determined by requiring the Maxwell potential to vanish at the horizon, 
\begin{equation}
  h(1)=1\Rightarrow \mu_{RN1}=Q,
\end{equation}
to avoid a singularity in the gauge field.
The speed of light can be normalized,
\begin{equation}
  c_1=1.
\end{equation} 
This leads to a single parameter $Q_1$ corresponding to the charge of the black hole, that defines the inner region.

This Q gives rise to a local chemical potential in the bulk, shown in figure \ref{EC:fig:RNchempot}. This potential always starts at $\mu(1)=0$ rises to a maximum, and decreases to $\mu(0)=0$, as previously stated.

We stress that this local chemical potential is not the whole story, because we wish our model to not only be pure AdS-RN. Eventually the local chemical potential will be large enough to support a fermion density, and we will no longer be in a purely AdS-RN solution. It does however give the location of the inner bound of the cloud.

\subsubsection{Region II: The Electron Cloud}

When working in the limit where the temperature of the cloud is zero, there is a well defined edge of the cloud, determined by the particle mass $m$ giving a region $z_1\geq z \geq z_2$ where the fluid is supported such that $\mu(z_1)=\mu(z_2)=m$. The inner bound $z_1$ is clearly given by the RN background solution, while $z_2$ has to be calculated numerically, for each value of $Q,m$ and $\beta$. The number density of cloud particles affects the bulk chemical potential, resulting in clouds of different size depending on the mass $m$ and the density parameter $\beta$.

In figure \ref{EC:fig:ECBounds}, the inner and outer bounds of the electron cloud for different $m$ are shown for $Q^2=4$ and $\beta=0,10,30$ and $100$. As the cloud particles have a greater charge than mass ($m<1$), their presence extends the region where the bulk chemical potential is larger than $m$. At $m=0.3$, having $\beta=10$ shifts the outer bound to $z=0.1$ from $z=0.2$ at $\beta=0$. This also gives rise to a critical limit, since a sufficiently dense electron cloud could source itself indefinitely, as can be seen in the figure by the absence of an outer bound at $\beta=30$ for masses below $0.46$. For each mass this critical limit can be written as $\beta<\beta_{crit}(m)$. Beyond that limit, the solutions are unphysical as they give rise to an infinitely dense and infinitely extended cloud in the bulk, and an infinite chemical potential and an infinite charge density on the conformal boundary. This is in contrast to the physically relevant setting where the charge density or chemical potential of a system is held fixed. The expansion of the cloud can then be interpreted, since the scale is set by the horizon radius, as the black hole shrinking, until the system ultimately cools down to the zero temperature solution, the electron star.

\subsubsection{Region III: Outer RN}
Matching the background RN-solution onto the outer surface is quite straightforward. It is however worth to note that the same four parameters now take on different values than previously, especially the speed of light which can no longer be normalized to 1. This has some implications to systems where one wishes to have different speeds of light in the IR and UV.

\subsection{Linear response}

\subsubsection{Region I: Inner RN}

In this region, having an explicit background solution and vanishing fluid quantities; pressure, number density and energy density, greatly simplifies the equations for the perturbations. These also lead to the elimination of the $\delta\!\phi$ equation, since all terms in that equation are proportional to either of those quantities or their derivatives.

The main issue here is the behaviour of the perturbations when approaching the horizon. This is handled by factoring out an exponential term in the fields and in the equations (by performing a Fröbenius-expansion). We are then left with a rather simple explicit set of equations that can all be solved numerically. After eliminating outgoing solutions, we find that there are two degrees of freedom left. These can be chosen to be $\delta\!A_x^*(1)$ and $\delta\!g_{xx}^*(1)$ (stars indicating post Fröbenius expansion). As we study linear differential equations, we can solve for two different sets of starting values, e.g.~(1,0) and (0,1), and eventually make linear combinations.

\subsubsection{Region II: The Electron Cloud}

The starting point of the next region is given by the previously obtained $z_1$. The matching at the inner surface is well defined and provides no further degrees of freedom, even though the number of fields and equations is increased.

\subsubsection{Region III: Outer RN}
Matching the perturbations is done similarly at $z_2$ as done at $z_1$. At the AdS boundary at $z=0$ is typically where you impose other conditions on the system, as these can describe physical quantities in the boundary theory, but it's easier to give different sets of starting values at the horizon, and then use the linearity of the system to find a suitable linear combination of solutions that also fulfills the boundary conditions \cite{Amado:2009ts}.

\subsubsection{Pure gauge solutions}
Working in radial gauge does not completely fix the gauge freedom of the system. Although the physical requirement of having infalling modes at the horizon does, that is not a requirement one necessarily wants to impose on pure gauge modes. The gauge freedoms can be written down in terms of four additional analytic solutions to the equations of motion
\begin{equation}
\begin{multicases}{2}
\delta\!g_{tt}(z)_3=0,\quad&
\delta\!g_{tx}(z)_3=0,\quad&
\delta\!g_{xx}(z)_3=0,\\
\delta\!g_{yy}(z)_3=0,\quad&
\delta\!A_{t}(z)_3= -i \omega,\quad&
\delta\!A_{x}(z)_3= i k, 
\end{multicases}
\end{equation}
\begin{equation}
\begin{multicases}{2}
\delta\!g_{tt}(z)_4=2 i \omega f(z),\quad&
\delta\!g_{tx}(z)_4=-i k f(z),\quad&
\delta\!g_{xx}(z)_4=0,\\
\delta\!g_{yy}(z)_4=0,\quad&
\delta\!A_{t}(z)_4=-i \omega h(z),\quad&
\delta\!A_{x}(z)_4=i k h(z),
\end{multicases}
\end{equation}
\begin{equation}
\begin{multicases}{2}
\delta\!g_{tt}(z)_5=0,\quad&
\delta\!g_{tx}(z)_5=-\frac{i \omega}{z^2},\quad&
\delta\!g_{xx}(z)_5=\frac{2 i k}{z^2},\\
\delta\!g_{yy}(z)_5=0,\quad&
\delta\!A_{t}(z)_5=0,\quad&
\delta\!A_{x}(z)_5=0,
\end{multicases}
\end{equation}
\begin{equation}
\begin{cases}
\delta\!g_{tt}(z)_6=i \left(2 \omega f(z) I_1+\frac{i f'(z)}{\sqrt{g(z)}}\right),\quad&
\delta\!g_{tx}(z)_6=-\frac{i \left(k z^2 f(z) I_1+\omega I_2\right)}{z^2},\\
\delta\!g_{xx}(z)_6=\frac{2 i \left(k z I_2+\frac{i}{\sqrt{g(z)}}\right)}{z^3},\quad&
\delta\!g_{yy}(z)_6=-\frac{2}{z^3 \sqrt{g(z)}},\\
\delta\!A_{t}(z)_6=-i \omega h(z) I_1-i \omega I_3+\frac{h'(z)}{\sqrt{g(z)}}\quad&
\delta\!A_{x}(z)_6=i k h(z) I_1+i k I_3,
\end{cases}
\end{equation}
where we have introduced the following integrals,
\begin{eqnarray}
    I_1&=&\int_1^{z} -\frac{i \omega \sqrt{g(z')}}{f(z')} \, dz'\,,\\
    I_2&=&\int_1^{z} -i k {z'}^2 \sqrt{g(z')} \, dz'\,,\\
    I_3&=&\int_1^{z} \frac{i \omega \sqrt{g(z')} h(z')}{f(z')} \, dz'\,.
\end{eqnarray}

\subsubsection{Boundary Conditions}
Due to the linearity of the differential equations, we can solve the system for any starting values at the horizon, get two different solutions and examine whether a linear combination of these, together with the pure gauge solutions, fulfill the boundary conditions on the AdS boundary. This is intuitively done by computing the determinant to see if it is zero \cite{Amado:2009ts}.

E.g.~to get the QNMs, choose $\delta\! A_x^*(z\!=\!1)=1$ and $\delta\! g_{tx}^*(z\!=\!1)=0$ to get a first set of solutions, choose $\delta\! A_x^*(z\!=\!1)=0$ and $\delta\! g_{tx}^*(z\!=\!1)=1$ to get a second. Together with the four pure gauge solutions, these make up a 6-by-6 matrix of boundary values. Then study the determinant
\begin{equation}
\left|\begin{array}{cccccc}
\delta\! g_{\text{tt}}(z)_{1} & \delta\! g_{\text{tx}}(z)_{1} & \delta\! g_{\text{xx}}(z)_{1} & \delta\! g_{\text{yy}}(z)_{1} & \delta\! A_{\text{t}}(z)_{1} & \delta\! A_{\text{x}}(z)_{1} \\
\delta\! g_{\text{tt}}(z)_{2} & \delta\! g_{\text{tx}}(z)_{2} & \delta\! g_{\text{xx}}(z)_{2} & \delta\! g_{\text{yy}}(z)_{2} & \delta\! A_{\text{t}}(z)_{2} & \delta\! A_{\text{x}}(z)_{2} \\
& &\cdots
\end{array}\right|_{z\to0},
\end{equation}
and find for which values of $(\omega,k)$ the determinant is zero. These pairs are points on a dispersion relation. Note that e.g.~$\omega(k)$ in general is complex for real $k$.

If one wants to study self-sourcing modes in a Maxwell boundary theory, one needs to adjust the boundary conditions accordingly.
This corresponds to the standard plasma-oscillation condition that the dielectric function, $\epsilon$, is zero, or that both the electric field and the current vanishes on the boundary, see e.g.~\cite{Aronsson:2017dgf,Aronsson:2018yhi}. In our holographic model, these is captured by the boundary condition
\begin{eqnarray}
\omega^2 \delta\! A_x+c\,e\, \delta\!A_x'&=&0\, .
\label{eq:plasmoncondtion}
\end{eqnarray}
where $e$ is the boundary Maxwell coupling. 
This is in addition to requiring $\delta\! g_{\mu\nu}=0$ at the AdS boundary, which would otherwise introduce dynamical gravity effects on the boundary and $\delta\!A_t=0$ at the boundary, which translates to keeping the chemical potential fixed.

\subsection{Equations for the Perturbations}\label{app:pert.eq}
\subsubsection{Equations of motion}
$\delta\!g_{tt}$-variation equation:
\begin{flalign*}
& \delta\!g_{tt}''
+2  f^{1/2} g \sigma \delta\!A_t
-  \frac{k  f^{3/2} g \sigma z^2}{\omega} \delta\!A_x
-  \frac{k  g \Bigl(\omega^2 + f^{1/2} h \sigma\Bigr) z^2}{\omega} \delta\!g_{tx}\\
&-  \frac{i  \Bigl(2 \omega^2 f^{1/2} g \sigma + k^2 f^{3/2} g \sigma z^2\Bigr)}{\omega} \delta\!\phi
+ \tfrac{1}{4}\bigl(-2 f z -  z^2 f'\bigr)  \delta\!g_{xx}' \\
&+ \tfrac{1}{4} \bigl(-2 f z -  z^2 f'\bigr) \delta\!g_{yy}' 
+\frac{\bigl(-2 f g - 2 g z f' -  f z g'\bigr)}{2 f g z} \delta\!g_{tt}' 
+ 3  h' \delta\!A_t'\\
&+ \frac{i f^{1/2} \bigl(-4 f g h \sigma + 2 g h \sigma z f' -  f h \sigma z g' - 2 f g \sigma z h' + 2 f g h z \sigma'\bigr)}{2 \omega g h z} \delta\!\phi'  \\
&+ \frac{i f^{3/2} \sigma }{\omega} \delta\!\phi''\\
&+ \frac{1}{2 f^2 g z^2}\Bigl(10 f^2 g - 6 f^2 g^2 z^2 + 2 f^2 g^2 \rho z^2 + 2 f^{3/2} g^2 h \sigma z^2 -  k^2 f^2 g^2 z^4 + 2 f g z f' \\
&\pushright{+ 2 g z^2 \bigl(f'\bigr)^2 + 2 f^2 z g' + f z^2 f' g' + 4 f g z^2 \bigl(h'\bigr)^2 - 2 f g z^2 f''\Bigr)\delta\!g_{tt} } \\
&+ \frac{1}{4 f g} \Bigl(4 f^2 g - 2 \omega^2 f g^2 z^2 - 12 f^2 g^2 z^2 + 4 f^2 g^2 \rho z^2 - 6 f^{3/2} g^2 h \sigma z^2 - 4 f g z f'\\
&\pushright{ -  g z^2 \bigl(f'\bigr)^2 + 2 f^2 z g' -  f z^2 f' g' - 2 f g z^2 \bigl(h'\bigr)^2 + 2 f g z^2 f''\Bigr)\delta\!g_{xx}} \\
&+ \frac{1}{4 f g}\Bigl(4 f^2 g - 2 \omega^2 f g^2 z^2 - 12 f^2 g^2 z^2 + 4 f^2 g^2 \rho z^2 - 6 f^{3/2} g^2 h \sigma z^2 - 2 k^2 f^2 g^2 z^4\\
&\pushright{ - 4 f g z f' -  g z^2 \bigl(f'\bigr)^2 + 2 f^2 z g' -  f z^2 f' g' - 2 f g z^2 \bigl(h'\bigr)^2 + 2 f g z^2 f''\Bigr)\delta\!g_{yy} }\\
&\pushright{=0}
\end{flalign*}

$\delta\!g_{tx}$-variation equation:
\begin{flalign*}
&\delta\!g_{tx}''
+2 \delta\!A_x f^{1/2} g \sigma 
+ 2i k \delta\!\phi f^{1/2} g \sigma 
+ k \omega \delta\!g_{yy} g z^2 
+ \frac{\delta\!g_{tx}' \bigl(- g f' -  f g'\bigr)}{2 f g}  \\
&+ 2 \delta\!A_x' h'
+ \delta\!g_{tx} \Bigl(-6 g + 2 g \rho + \frac{4}{z^2} -  \frac{f'}{f z} + \frac{g'}{g z} + \frac{\bigl(h'\bigr)^2}{f}\Bigr) =0
\end{flalign*}

$\delta\!g_{xx}$-variation equation:
\begin{flalign*}
&+ \delta\!g_{xx}''
- \frac{k \delta\!A_x f^{1/2} g \sigma}{\omega} 
-  \frac{i k^2 \delta\!\phi f^{1/2} g \sigma}{\omega} 
+ \frac{k \delta\!g_{tx} g \Bigl(\omega^2 -  f^{1/2} h \sigma\Bigr)}{\omega f} \\
&+ \frac{\delta\!g_{yy}' \bigl(-2 f -  z f'\bigr)}{4 f z} 
+ \frac{\delta\!g_{xx}' \bigl(6 f g + g z f' - 2 f z g'\bigr)}{4 f g z} 
+ \frac{\delta\!A_t' h'}{f z^2} \\
&+ \frac{\delta\!g_{tt} \Bigl(10 f g - 6 f g^2 z^2 + 2 f g^2 \rho z^2 + k^2 f g^2 z^4 + 2 f z g' + 2 g z^2 \bigl(h'\bigr)^2\Bigr)}{2 f^2 g z^4} \\
&+ \frac{i \delta\!\phi' \bigl(-4 f g h \sigma + 2 g h \sigma z f' -  f h \sigma z g' - 2 f g \sigma z h' + 2 f g h z \sigma'\bigr)}{2 \omega f^{1/2} g h z^3} \\
&+ \frac{i f^{1/2} \sigma \delta\!\phi''}{\omega z^2} \\
&+ \frac{1}{4 f^2 g z^2} \Bigl(-12 f^2 g - 2 \omega^2 f g^2 z^2 + 12 f^2 g^2 z^2 - 4 f^2 g^2 \rho z^2 + 2 f^{3/2} g^2 h \sigma z^2 \\
&\pushright{- 2 k^2 f^2 g^2 z^4 + g z^2 \bigl(f'\bigr)^2 - 2 f^2 z g' + f z^2 f' g' + 2 f g z^2 \bigl(h'\bigr)^2 - 2 f g z^2 f''\Bigr)\delta\!g_{yy}} \\
&+ \frac{1}{4 f^2 g z^2}\Bigl(-4 f^2 g + 2 \omega^2 f g^2 z^2 - 12 f^2 g^2 z^2 + 4 f^2 g^2 \rho z^2 - 6 f^{3/2} g^2 h \sigma z^2 \\
&\pushright{-  g z^2 \bigl(f'\bigr)^2 - 2 f^2 z g' -  f z^2 f' g' - 2 f g z^2 \bigl(h'\bigr)^2 + 2 f g z^2 f''\Bigr)\delta\!g_{xx} }\\
&\pushright{=0}
\end{flalign*}

$\delta\!g_{yy}$-variation equation:
\begin{flalign*}
& \delta\!g_{yy}'' 
- \frac{k \delta\!A_x f^{1/2} g \sigma}{\omega} 
-  \frac{i k^2 \delta\!\phi f^{1/2} g \sigma}{\omega} 
-  \frac{k \delta\!g_{tx} g \Bigl(\omega^2 + f^{1/2} h \sigma\Bigr)}{\omega f} \\
&+ \frac{\delta\!g_{xx}' \bigl(-2 f -  z f'\bigr)}{4 f z} 
+ \frac{\delta\!g_{yy}' \bigl(6 f g + g z f' - 2 f z g'\bigr)}{4 f g z} 
+ \frac{\delta\!A_t' h'}{f z^2} \\
&+ \frac{\delta\!g_{tt} \Bigl(10 f g - 6 f g^2 z^2 + 2 f g^2 \rho z^2 -  k^2 f g^2 z^4 + 2 f z g' + 2 g z^2 \bigl(h'\bigr)^2\Bigr)}{2 f^2 g z^4} \\
&+ \frac{i \delta\!\phi' \bigl(-4 f g h \sigma + 2 g h \sigma z f' -  f h \sigma z g' - 2 f g \sigma z h' + 2 f g h z \sigma'\bigr)}{2 \omega f^{1/2} g h z^3} 
+ \frac{i f^{1/2} \sigma \delta\!\phi''}{\omega z^2} \\
&+ \frac{1}{4 f^2 g z^2} \Bigl(-12 f^2 g - 2 \omega^2 f g^2 z^2 + 12 f^2 g^2 z^2 - 4 f^2 g^2 \rho z^2 + 2 f^{3/2} g^2 h \sigma z^2 \\
&\pushright{+ g z^2 \bigl(f'\bigr)^2 - 2 f^2 z g' + f z^2 f' g' + 2 f g z^2 \bigl(h'\bigr)^2 - 2 f g z^2 f''\Bigr)\delta\!g_{xx}} \\
&+ \frac{1}{4 f^2 g z^2} \Bigl(-4 f^2 g + 2 \omega^2 f g^2 z^2 - 12 f^2 g^2 z^2 + 4 f^2 g^2 \rho z^2 - 6 f^{3/2} g^2 h \sigma z^2 \\
&\pushright{- 2 k^2 f^2 g^2 z^4 -  g z^2 \bigl(f'\bigr)^2 - 2 f^2 z g' -  f z^2 f' g' - 2 f g z^2 \bigl(h'\bigr)^2 + 2 f g z^2 f''\Bigr)\delta\!g_{yy}}\\
&\pushright{=0}
\end{flalign*}

$\delta\!A_{t}$-variation equation:
\begin{flalign*}
& \delta\!A_t''  
- k^2 \delta\!A_t g z^2 
+ \frac{k \delta\!g_{tx} f^{1/2} g \sigma z^2}{\omega} 
+ \frac{i k^2 \delta\!\phi f^{3/2} g \sigma z^2}{\omega h} \\
&+ \frac{\delta\!A_x \Bigl(- k \omega^2 g h z^2 + k f^{3/2} g \sigma z^2\Bigr)}{\omega h} 
+ \frac{\delta\!A_t' \bigl(-4 f g -  g z f' -  f z g'\bigr)}{2 f g z} 
+ \frac{\delta\!g_{tt}' h'}{2 f} \\
&+ \tfrac{1}{2} z^2 \delta\!g_{xx}' h' 
+ \tfrac{1}{2} z^2 \delta\!g_{yy}' h' 
+ \tfrac{1}{2} \delta\!g_{xx} \Bigl(f^{1/2} g \sigma z^2 + 2 z h'\Bigr) 
+ \tfrac{1}{2} \delta\!g_{yy} \Bigl(f^{1/2} g \sigma z^2 + 2 z h'\Bigr) \\
&-  \frac{i f^{1/2} \delta\!\phi' \bigl(-4 f g h \sigma + 2 g h \sigma z f' -  f h \sigma z g' - 2 f g \sigma z h' + 2 f g h z \sigma'\bigr)}{2 \omega g h^2 z} \\
&-  \frac{i f^{3/2} \sigma \delta\!\phi''}{\omega h}
+ \frac{\delta\!g_{tt} \Bigl(- f^{3/2} g^2 \sigma z - 4 f g h' - 2 g z f' h' -  f z g' h' + 2 f g z h''\Bigr)}{2 f^2 g z}=0
\end{flalign*}

$\delta\!A_{x}$-variation equation:
\begin{flalign*}
&\frac{k \omega \delta\!A_t g}{f} 
-  \frac{i k \delta\!\phi f^{1/2} g \sigma}{h} 
+ \frac{\delta\!A_x g \Bigl(\omega^2 h 
-  f^{3/2} \sigma\Bigr)}{f h} 
+ \frac{\delta\!A_x' \bigl(g f' -  f g'\bigr)}{2 f g} 
+ \frac{\delta\!g_{tx}' h'}{f}  \\
&+ \delta\!A_x''
+ \frac{\delta\!g_{tx} \Bigl(-2 f^{3/2} g^2 \sigma -  g f' h' -  f g' h' + 2 f g h''\Bigr)}{2 f^2 g}=0
\end{flalign*}

$\delta\!\phi$-equation, from $n$-variation:
\begin{flalign*}
& \delta\!\phi''
+ i k \delta\!A_x g z^2 
+ \frac{i k \delta\!g_{tx} g h z^2}{f} 
+ \frac{i \omega \delta\!g_{xx} g h z^2}{2 f} 
+ \frac{i \omega \delta\!g_{yy} g h z^2}{2 f} 
+ \frac{i \omega \delta\!A_t g h \sigma'}{f^{3/2} \sigma \mu'} \\
&+ \frac{i \omega \delta\!g_{tt} g h^2 \sigma'}{2 f^{5/2} \sigma \mu'} 
-  \tfrac{1}{2} \delta\!\phi' \bigl(\frac{4}{z} -  \frac{2 f'}{f} + \frac{g'}{g} + \frac{2 h'}{h} -  \frac{2 \sigma'}{\sigma}\bigr) 
-  \delta \phi g \Bigl(k^2 z^2 -  \frac{\omega^2 h \sigma'}{f^{3/2} \sigma \mu'}\Bigr)=0 
\end{flalign*}

\subsubsection{Constraint eqations}

$\delta\!g_{tz}$-variation, constraint equation:
\begin{flalign*}
&\frac{i k z^2 \delta\!g_{tx}'}{2 f g} 
+ \frac{i \omega z^2 \delta\!g_{xx}'}{2 f g} 
+ \frac{i \omega z^2 \delta\!g_{yy}'}{2 f g} 
-  \frac{\sigma \delta\!\phi'}{f^{1/2} g} 
-  \frac{i k \delta\!g_{tx} z^2 f'}{2 f^2 g} 
+ \frac{i \omega \delta\!g_{xx} z \bigl(2 f -  z f'\bigr)}{4 f^2 g} \\
&+ \frac{i \omega \delta\!g_{yy} z \bigl(2 f -  z f'\bigr)}{4 f^2 g}=0
\end{flalign*}

$\delta\!g_{xz}$-variation, constraint equation:
\begin{flalign*}
&- \frac{i \omega \delta\!g_{tx} z}{f g} 
+ \frac{i k \delta\!g_{yy} z^3}{g} 
-  \frac{i k z^2 \delta\!g_{tt}'}{2 f g} 
-  \frac{i \omega z^2 \delta\!g_{tx}'}{2 f g}
+ \frac{i k z^4 \delta\!g_{yy}'}{2 g} \\
&+ \frac{i \delta\!g_{tt} z \bigl(-2 k f + k z f'\bigr)}{4 f^2 g} 
-  \frac{i k \delta\!A_t z^2 h'}{f g} 
-  \frac{i \omega \delta\!A_x z^2 h'}{f g}=0
\end{flalign*}

$\delta\!g_{zz}$-variation, constraint equation:
\begin{flalign*}
&\frac{\delta\!A_t \sigma}{f^{1/2} g} 
-  \frac{i \omega \delta\!\phi \sigma}{f^{1/2} g} 
-  \frac{k \omega \delta\!g_{tx} z^2}{f g} 
-  \frac{\delta\!g_{tt}'}{f g^2 z} 
+ \frac{z \delta\!g_{xx}' \bigl(2 f -  z f'\bigr)}{4 f g^2} 
+ \frac{z \delta\!g_{yy}' \bigl(2 f -  z f'\bigr)}{4 f g^2} \\
&+ \frac{\delta\!g_{xx} \Bigl(2 f -  z \bigl(\omega^2 g z + f'\bigr)\Bigr)}{2 f g^2} 
+ \frac{\delta\!g_{yy} \Bigl(f \bigl(2 + k^2 g z^4\bigr) -  z \bigl(\omega^2 g z + f'\bigr)\Bigr)}{2 f g^2} \\
&-  \frac{\delta\!A_t' h'}{f g^2} 
+ \frac{\delta\!g_{tt} \Bigl(f^{1/2} g h \sigma z -  k^2 f g z^3 + 2 f' -  z \bigl(h'\bigr)^2\Bigr)}{2 f^2 g^2 z}=0
\end{flalign*}

$\delta\!A_{z}$-variation, constraint equation:
\begin{flalign*}
&\frac{i \omega \delta\!A_t'}{f g} 
+ \frac{i k z^2 \delta\!A_x'}{g} 
+ \frac{f^{1/2} \sigma \delta\!\phi'}{g h} 
+ \frac{i \omega \delta\!g_{tt} h'}{2 f^2 g} 
+ \frac{i k \delta\!g_{tx} z^2 h'}{f g} 
+ \frac{i \omega \delta\!g_{xx} z^2 h'}{2 f g} \\
&+ \frac{i \omega \delta\!g_{yy} z^2 h'}{2 f g}
\hfill=0
\end{flalign*}

\newpage

\bigskip

\bibliographystyle{JHEP}
\bibliography{ElectronCloud}

\end{document}